\documentclass[dvips,12pt]{article}
\usepackage{amsmath}
\usepackage{amsfonts}
\usepackage{amsthm}
\usepackage{amssymb}
\usepackage{amstext}
\usepackage{amsopn}
\usepackage{mathrsfs}
\usepackage{subfigure}
\usepackage{graphicx}
\usepackage[left=2.3cm,top=2cm,right=2.3cm,bottom=2cm, nohead,foot=1cm]{geometry}
\numberwithin{equation}{section}

\newtheorem{theorem}{Theorem}[section]
\newtheorem{lemma}[theorem]{Lemma}

\newtheorem{proposition}[theorem]{Proposition}
\newtheorem{definition}[theorem]{Definition}

\newtheorem{example}[theorem]{Example}

\newcommand{\R}{{\mathbb R}}

\newcommand{\Hi}{{\mathcal{H}}}
\newcommand{\Bi}{{B}}

\newcommand{\N}{{\mathbb N}}
\newcommand{\Tr}{{\textup{Tr}}}

\newcommand{\TC}{{\textup{T}}}

\newcommand{\C}{{\mathbb C}}

\title{Decoherence Rates for Galilean Covariant Dynamics }
\author{\textbf{Jeremy Clark}\\ jeremy@math.ucdavis.edu \\ University of California, Davis\\ One Shields Ave, Davis, CA 95616 }

\begin{document}
\maketitle

\begin{abstract}
We introduce a measure of decoherence for a class of density operators.  For Gaussian density operators in dimension one it coincides with an index used by Morikawa (1990).   Spatial decoherence rates are derived for three large classes of the Galilean covariant quantum semigroups introduced by Holevo.   We also characterize the relaxation to a Gaussian state for these dynamics and  give a theorem for the convergence of the Wigner function to the probability distribution of the classical analog of the process.   
  
\end{abstract}

\section{Introduction}

One important phenomenon in quantum optics is the suppression of wave behavior for a quantum particle interacting with an environment.   This subdued wave behavior is usually referred to as decoherence and is strongly emphasized by many physicists~\cite{Decoh} as being a major ingredient for the construction of a macroscopic world that is well-approximated by models of localized objects following well-defined trajectories.    Apart from the natural theoretical appeal of this topic,  quantifying spatial decoherence has also attracted interest from experimental physicists working in quantum optics~\cite{Exper, Talbot}.   
 
If $\rho_{t}$ is the reduced density operator of a particle with spatial degrees of freedom at time $t$ interacting with a environment, then the rough intuition is that the particle is undergoing spatial decoherence if the off-diagonal position ket entries $x_{1}\neq x_{2}$ of $\rho_{t}(x_{1},x_{2})$ vanish at exponential rates.   Thus the particle decohering through an environmental interaction is in some sense becoming more diagonal in the $x$-basis.   In the present paper, we study certain categories of  dynamics for decoherence by introducing a coherence index of the form:
\begin{align}\label{Commutivity} 
S_{\vec{X}}(\rho)=\frac{C_{\vec{X}}(\rho) }{D_{\vec{X}}(\rho)}= \frac{ \Big(\frac{1}{\|\rho\|_{2}^{2}}  \sum_{j=1}^{d}\Tr[-[X_{j},\rho]^{2}]  \Big)^{\frac{1}{2}} }{  \Big( \frac{1}{\|\rho\|_{2}^{2}}\sum_{j=1}^{d}\Tr[\{ X_{j}-\Tr[X_{j}\rho] ,\rho \}^{2} ]  \Big)^{\frac{1}{2}}   },
\end{align}
for a density operator $\rho$, where  $X_{j}$ $j=1,\cdots d$ are the position operators for a particle traveling with $d$ spatial degrees of freedom.    The numerator $C_{\vec{X}}(\rho)$ is a coherence length-like quantity while the denominator $D_{\vec{X}}(\rho)$ is a standard deviation-like quantity.  
    We study the above index for a density operator $\Gamma_{t}(\rho)$ in the limit $t\rightarrow \infty$, where $\Gamma_{t}$ is a dynamical semigroup of trace preserving maps formally satisfying the equation: 
\begin{multline}\label{NoSymm}
\frac{d}{dt}\Gamma_{t}(\rho)=i[|\vec{K}|^{2},\Gamma_{t}(\rho)]-\frac{1}{2}\sum_{i,j}A^{x,x}_{i,j}[X_{i},[X_{j},\Gamma_{t}(\rho) ]]-\sum_{i,j}A^{x,k}_{i,j}[X_{i},[K_{j},\Gamma_{t}(\rho)]]\\ -\frac{1}{2}\sum_{i,j}A^{k,k}_{i,j}[K_{i},[K_{j},\Gamma(\rho)]] +\int d\mu(\mathbf{x},\mathbf{k})(W_{\mathbf{x},\mathbf{k}}^{*}\Gamma_{t}(\rho)W_{\mathbf{x},\mathbf{k}}-\Gamma_{t}(\rho) ).
\end{multline}
In the above, $\vec{K}$ is the vector of momentum operators, $X_{j}$ for $j=1,\dots d$ are the position operators, $W_{(\mathbf{x},\mathbf{k})}=e^{i\mathbf{k}\vec{X}+i\mathbf{x}\vec{K}} $ is the Weyl operator corresponding to translation in phase space by $(\vec{q},\vec{p})$,  $\mu$ is a symmetric measure about the origin on $\R^{d}\times \R^{d}$ satisfying $\int d\mu(\mathbf{x},\mathbf{k})(|\mathbf{x}|^{2}+|\mathbf{k}|^{2})<\infty$, and  $A^{x,x}, (A^{x,k})^{t}=A^{k,x}, A^{k,k}$ are  the $d\times d$ block matrices of  a positive semidefinite real valued matrix $A$:   
$$ A=\begin{pmatrix} A^{x,x} & A^{x,k}\\ A^{k,x} & A^{k,k} \end{pmatrix}. $$
The dynamics $\Gamma_{t}$ describes a free particle (no forcefield potential) in a random environment giving the particle a L\'evy process of phase space kicks through  conjugation by the Weyl operators.    The quadratic terms in $[X_{j},\cdot]$ and $[K_{j},\cdot]$ correspond to a continuous limit of frequent small kicks.  Later in this introduction, this model and related models will be discussed further.    
 
 Define the $2d\times 2d$ matrix: $B=\int d\mu(\mathbf{x},\mathbf{k}) \begin{tiny}\begin{pmatrix} \mathbf{x} \\ \mathbf{k} \end{pmatrix}\end{tiny}\otimes \begin{tiny}\begin{pmatrix} \mathbf{x} \\ \mathbf{k} \end{pmatrix}\end{tiny}   $.  Let $B^{x,x},B^{x,k},B^{k,x},B^{k,k}$ be the blocks of $B$: $ \begin{tiny} \begin{pmatrix} B^{x,x} & B^{x,k}\\ B^{k,x} & B^{k,k} \end{pmatrix}\end{tiny} $.  The analysis of the asymptotics of $S_{\vec{X}}(\Gamma_{t}(\rho))$  splits into three main categories.  Let $\nu$ be some positive measure on $\R^{d}$.  

\begin{enumerate}
 \item  \label{one} Only jumps in momentum: $A^{x,k}=A^{k,x}=A^{k,k}=0$ and $\mu(\mathbf{x},\mathbf{k})=\delta(\mathbf{x})\nu(\mathbf{k})$, where   $A^{x,x}$  is assumed to be positive definite or $\nu$ is assumed to have a density.   
\begin{align} \label{MomOnly}S_{\vec{X}}(\Gamma_{t}(\rho))\sim t^{-2}\sqrt{3}\frac{\Tr[(A^{x,x}+B^{x,x})^{-1}]^{\frac{1}{2}}}{\Tr[A^{x,x}+B^{x,x}]^{\frac{1}{2}}} 
\end{align}

\item \label{two} Only jumps in position: $A^{x,x}=A^{x,k}=A^{k,x}=0$ and $\mu(\mathbf{x},\mathbf{k})=\nu(\mathbf{x})\delta(\mathbf{k})$, where $A^{k,k}$  is assumed to be positive definite or $\nu$ is assumed to have a density.  
\begin{align}\label{PosOnly} S_{\vec{X}}(\Gamma_{t}(\rho))\sim  t^{-\frac{1}{2} }2^{\frac{1}{2}} \Tr[(A^{k,k})^{-1}]^{\frac{1}{2}} \frac{ (  \int d\mathbf{k}|\rho(\mathbf{k},\mathbf{k})|^{2} | \mathbf{k}-E[\vec{K}\rho]|^{2} )^{\frac{1}{2}}}{(\int d\mathbf{k}|\rho(\mathbf{k},\mathbf{k})|^{2})^{\frac{1}{2}} } 
\end{align}

\item \label{three} Active presence of both jumps in momentum and position where $A$ is assumed to be positive definite or $\mu$ is assumed to have a density.   
\begin{align}\label{MomPos} S_{\vec{X}}(\Gamma_{t}(\rho))\sim t^{-2}\sqrt{3}\frac{\Tr[(A^{x,x}+B^{x,x})^{-1}]^{\frac{1}{2}}}{\Tr[A^{x,x}+B^{x,x}]^{\frac{1}{2}}}
\end{align}
 \end{enumerate}
 
For two functions $\alpha_{t}, \beta_{t}$, by $\alpha_{t}\sim \beta_{t}$ we mean that $\lim_{t\rightarrow \infty}\frac{\alpha_{t}}{\beta_{t}}=1$.   It is expected that the asymptotics will have an error on the order of $\mathit{O}(t^{-\frac{5}{2}})$ for cases~(\ref{one}) and~(\ref{three}) and $\mathit{O}(t^{-1})$ for case~(\ref{two}) due to the application of variations of Laplace's method in the approximations.     Similar statements can made for decoherence rates in momentum.  

Theorem~(\ref{Relaxation}) characterizes the relaxation for the dynamics $\Gamma_{t}$ in the cases $(1)$ and $(3)$ above.    It is essentially a central limit theorem where the Poisson noise can be approximated by a Gaussian noise and all information from the initial state is lost.    For dimension one, case $(1)$ and certain generalizations are discussed in~\cite{Lutz} and~\cite{Levy}.   In the current article, we characterize the relaxation as a Hilbert-Schmidt norm convergence of the process $\Gamma_{t}(\rho)$ for an initial $\rho$ to a process $\tilde{\rho}_{t}$ defined as
  \begin{align}\label{Rel}
  \tilde{\rho}_{t}(\mathbf{x}_{1},\mathbf{x}_{2})=\frac{(\det(c_{t}))^{\frac{1}{2}} }{(2\pi)^{\frac{d}{2}} } e^{-\frac{1}{4}\langle \mathbf{x}_{1}-\mathbf{x}_{2}|\mathbf{a}_{t}(\mathbf{x}_{1}-\mathbf{x}_{2})\rangle-\frac{i}{2}\langle \mathbf{x}_{1}-\mathbf{x}_{2}|\mathbf{b}_{t}(\mathbf{x}_{1}+\mathbf{x}_{2})\rangle-\frac{1}{4}\langle \mathbf{x}_{1}+\mathbf{x}_{2}|\mathbf{c}_{t}(\mathbf{x}_{1}+\mathbf{x}_{2})\rangle  },
  \end{align}
 where $\mathbf{a}_{t}=t(A^{x,x}+B^{x,x})$, $\mathbf{b}_{t}=\frac{3}{t}I$, and $\mathbf{c}_{t}=\frac{3}{t^{3}}(A^{x,x}+B^{x,x})^{-1}$.   Since $\Gamma_{t}(\rho)$ and $\tilde{\rho}_{t}$ tend to zero in $\|\cdot\|_{2}$, in~(\ref{Relaxation}) we characterize the convergence for $t\rightarrow \infty$ as
 \begin{align}\label{Convergence}
 \frac{\|\Gamma_{t}(\rho)-\tilde{\rho}_{t}\|_{2}}{ \|\tilde{\rho}_{t}\|_{2}}\rightarrow 0.
 \end{align}
 As discussed in~\cite{Lutz} and~\cite{Levy}, non-Gaussian limiting dynamics corresponding to more general stable laws can emerge if the L\'evy measure $\nu$ in case $(1)$ does not have finite second moments.   
 
 Also for cases $(1)$ and $(3)$, Theorem~(\ref{Classicality}) characterizes the convergence of the Wigner distributions $\mathcal{W}_{\Gamma_{t}(\rho)}(\mathbf{x},\mathbf{v})$ to the probability distributions of the classical counterpart of the dynamics.    Let $(\tilde{x}_{t},\tilde{k}_{t})\in \R^{2n}$ be the stochastic process in phase space with characteristic functions
 $$\varphi_{(\tilde{x}_{t},\tilde{k}_{t})}(\mathbf{q},\mathbf{p})= e^{t \mathit{l}(\mathbf{q},\mathbf{p})},   $$
 where
$$\mathit{l}(\mathbf{q},\mathbf{p})= -\frac{1}{2}\langle \mathbf{p}|A^{x,x}\mathbf{p}\rangle-\langle \mathbf{p}|A^{x,k}\mathbf{q}\rangle-\frac{1}{2}\langle \mathbf{p}|A^{k,k}\mathbf{p}\rangle + \int d\mu(\mathbf{x},\mathbf{k})(e^{i\mathbf{p}\cdot \mathbf{x}+i\mathbf{q}\cdot \mathbf{k}}-1).$$
 The stochastic process $(\tilde{x}_{t}+\int_{0}^{t} ds \tilde{k}_{s} ,\tilde{k}_{t})$ describes a classical particle initially at the origin with zero momentum undergoing free evolution interrupted by phase space jumps determined by the L\'evy process  $(\tilde{x}_{t},\tilde{k}_{t})$.    Let $p_{t}(\mathbf{x},\mathbf{v})$ be the probability distribution at time $t$ for the process $(\tilde{x}_{t}+\int_{0}^{t} ds \tilde{k}_{s} ,\tilde{k}_{t})$, then~(\ref{Classicality}) states that
 \begin{align}\label{Class}
 \frac{\big(\int d\mathbf{x}d\mathbf{v}| \mathcal{W}_{\Gamma_{t}(\rho)}(\mathbf{x},\mathbf{v})-p_{t}(\mathbf{x},\mathbf{v})|^{2} \big)^{\frac{1}{2}}}{\big(\int d\mathbf{x}d\mathbf{v}| p_{t}(\mathbf{x},\mathbf{v})|^{2} \big)^{\frac{1}{2}}}\rightarrow 0.
 \end{align}   
    
The index $S_{\vec{X}}(\rho)$ is intended as a measure of the spatial coherence of the density operator $\rho$.    In terms of the integral kernel of $\rho$ in the $\vec{X}$ basis, $S_{\vec{X}}(\rho)$ can be written:
\begin{align}\label{PosIndex}
S_{\vec{X}}(\rho)=\frac{ \Big( \int d\mathbf{x}_{1}d\mathbf{x}_{2}|\mathbf{x}_{1}-\mathbf{x}_{2}|^{2}|\rho(\mathbf{x}_{1},\mathbf{x}_{2})|^{2}\Big)^{\frac{1}{2}}         }{\Big( \int d\mathbf{x}_{1}d\mathbf{x}_{2}| \mathbf{x}_{1}+\mathbf{x}_{2}-2\vec{m} |^{2}|\rho(\mathbf{x}_{1},\mathbf{x}_{2})|^{2}\Big)^{\frac{1}{2}}        }\text{ , for } \Tr[\vec{X}\rho]=\vec{m}.    
\end{align}
The numerator measures a width of the integral kernel $\rho(\mathbf{x}_{1},\mathbf{x}_{2})$ in the off-diagonal direction while the denominator measures the width in the diagonal direction.     Since the diagonal of the kernel $\rho(\mathbf{x},\mathbf{x})$ has the interpretation as the probability density of  finding the particle at the point $\mathbf{x}$, the computation of the denominator is centered by the mean value $\Tr[\vec{X}\rho ]$.    $S_{\vec{X}}(\rho)$ takes values in the interval $(0,1]$ and is one in the case that $\rho$ is a pure state.    For the case when $\rho$ has a Gaussian form in one dimension
$$ \rho=\frac{2\sqrt{C}}{\sqrt{\pi}}e^{-A(x_{1}-x_{2})^{2}-iB(x_{1}^{2}-x_{2}^{2})-C(x_{1}+x_{2})^{2}-iD(x_{1}-x_{2})-E(x_{1}+x_{2})-F},$$
where all constants $A,\cdots, F$ are real, $A\geq C>0$, and $F=\frac{E^{2}}{4C}$, then  $S_{\vec{X}}(\rho)=\sqrt{\frac{C}{A}}$.       This agrees with the index used in~\cite{Morikawa} for decoherence models  where the Gaussian form is preserved.   A similar quantity coherence for momentum $S_{\vec{K}}(\rho)$ can also be defined.  

We are interested in the power laws that  arise in the study of $S_{\vec{X}}(\rho)$ when $\rho=\rho_{t}$ evolves according to some Markovian dynamics.    In the case where $\rho_{t}$ evolves as a free particle without noise (i.e. equation~(\ref{NoSymm}) with zero noise terms), then $S_{\vec{X}}(\rho_{t})$ will approach a non-zero constant.    We see from the difference between~(\ref{MomOnly}) and~(\ref{PosOnly}) that the power law for $S_{\vec{X}}(\rho_{t})$ will depend on the nature of the noise.   

There are bounds involving the numerator of $S_{\vec{X}}(\rho)$ and the denominator of $S_{\vec{K}}(\rho)$ and vice versa, since using the identities 
$$\rho= -i[X_{j},\{K_{j},\rho\}]+\{K_{j},i[X_{j},\rho]\} \text{ and } \rho=i[K_{j},\{ X_{j},\rho\}]-\{X_{j}, i[K_{j},\rho]\}, \text{ then }$$
\begin{align}\label{FunLim}
C_{\vec{X}}(\rho)D_{\vec{K}}(\rho)\geq \frac{1}{2} \text{ and } C_{\vec{K}}(\rho)D_{\vec{X}}(\rho)\geq \frac{1}{2}.
\end{align}
 The inequalities~(\ref{FunLim}) reduce to the uncertainty principle in the case where $\rho$ is a pure state, and they represent a fundamental lower bound on the coherence distance in position and the statistical spread in momentum in terms of the spread in momentum and position respectively.   The coherence length may only vanish as the spread in momentum grows without bound.   For more realistic dynamics~\cite{Vacchini0} including the effects of friction, the wavelength will tend to settle down to some finite value; the suppression of wave effects will depend on the wavelength  being small compared to relevant length scales.     
 
A more general understanding of the interplay between coherence length and momentum distribution can be found by looking at expression for $C_{\vec{X}}(\rho)=\|\rho\|_{2}^{-1}\| [\vec{X},\rho]\|_{2}$.  Since
  $$\nabla_{\mathbf{k}}(e^{i\mathbf{k}\vec{X}}\rho e^{-i\mathbf{k}\vec{X}})|_{\mathbf{k}=0}=i[\vec{X},\rho],$$
$i[\vec{X},\rho]$ gives the rate of change of the density operator $\rho$ given a small momentum shift in momentum $\mathbf{k}$.    $C_{\vec{X}}(\rho)$ is thus susceptible to being small in situations where the momentum distribution of $\rho$ is widely spread out and smooth.

    Unfortunately, the measure of the momentum spread appearing in~(\ref{FunLim}) is not simply the standard deviation: $(\Tr[\rho(\vec{K}-\Tr[\vec{K}\rho])^{2}])^{\frac{1}{2}}$ or a quantity naturally bounded by it.     We choose the quantity $D_{\vec{X}}(\rho)$ for our coherence index $S_{\vec{X}}(\rho)$, since  it bounds $D_{\vec{X}}(\rho)$ and satisfies~(\ref{FunLim}).   We could have also considered alternative indices
$$\frac{\|[\vec{X},\sqrt{\rho}]\|_{2}}{2\big(\Tr[\rho(\vec{X}-\Tr[\vec{X}\rho])^{2}]\big)^{\frac{1}{2}} }\text{ and }\frac{\|[\vec{K},\sqrt{\rho}]\|_{2}}{2\big(\Tr[\rho(\vec{K}-\Tr[\vec{K}\rho])^{2}]\big)^{\frac{1}{2}} },   $$
which  take on values in $[0,1]$ and have the same numerator-denominator inequalities as in~(\ref{FunLim}) hold.    The advantage of these quantities is that the denominator is actually a variance formula.   However, the numerator appears difficult to work with.

The convergence~(\ref{Convergence}) gives a detailed form of the limiting dynamics.   We can for instance write down the form of the density operator $\tilde{\rho}_{t}$ in the momentum basis: 
 \begin{align}\label{Momentum}
  \tilde{\rho}_{t}(\mathbf{k}_{1},\mathbf{k}_{2})=\frac{(\det(c_{t}^{\prime}))^{\frac{1}{2}} }{(2\pi)^{\frac{d}{2}} } e^{-\frac{1}{4}\langle \mathbf{k}_{1}-\mathbf{k}_{2}|\mathbf{a}^{\prime}_{t}(\mathbf{k}_{1}-\mathbf{k}_{2})\rangle-\frac{i}{2}\langle \mathbf{k}_{1}-\mathbf{k}_{2}|\mathbf{b}_{t}^{\prime}(\mathbf{k}_{1}+\mathbf{k}_{2})\rangle-\frac{1}{4}\langle \mathbf{k}_{1}+\mathbf{k}_{2}|\mathbf{c}_{t}^{\prime}(\mathbf{k}_{1}+\mathbf{k}_{2})\rangle  },
\end{align}
where $\mathbf{a}^{\prime}_{t}=\frac{t^{3}}{12}(A^{x,x}+B^{x,x})$, $\mathbf{b}^{\prime}_{t}=-\frac{t}{4}I$, and $\mathbf{c}^{\prime}_{t}=\frac{1}{4t}(A^{x,x}+B^{x,x})^{-1}$.   Since the eigenvalues of $\mathbf{c}^{\prime}_{t}$ are vanishing as $t^{-1}$, the average kinetic energy of the particle is growing linearly with $t$.    This corresponds to a high reservoir temperature assumption appearing in derivations of various subclasses of dynamics satisfying~(\ref{NoSymm}).    The off-diagonal width of $\tilde{\rho}_{t}(\mathbf{k}_{1},\mathbf{k}_{2})$ is on the order of $t^{-\frac{3}{2}}$.  With the naive identification $\mathbf{k}_{1}\approx \mathbf{k}_{2}$, ~(\ref{Momentum}) is like a Gibbs state $N_{t}^{-1}e^{-\beta_{t}|\vec{K}|^{2} },$
where $\beta_{t}=\frac{1}{4t}c$ (, where we have assumed that the environment is rotationally symmetric so that $A^{x,x}+B^{x,x}=cI$ for some constant $c$).    Thus the ``temperature" increases linearly in time.  
 As a consequence of the Gaussian form of $\tilde{\rho}_{t}$
$$S_{\vec{K}}(\tilde{\rho}_{t})=S_{\vec{X}}(\tilde{\rho}_{t}), $$
so the particle decoheres in momentum at the same law that it decoheres in position.   

~(\ref{Class}) gives another way to define the classicality of the system as $t\rightarrow \infty$ other than the vanishing of the off diagonal terms for the density matrix $\rho_{\Gamma_{t}(\rho)}$ in the position and the momentum basis.    By taking the Fourier transform of $\tilde{\rho}_{t}(\mathbf{x}_{1},\mathbf{x}_{2})$ in the $\mathbf{x}_{1}-\mathbf{x}_{2}$ direction we can obtain the relaxation form for the Wigner function $\mathcal{W}_{\Gamma_{t}(\rho)}(\mathbf{x},\mathbf{v})$ and the classical process $p_{t}(\mathbf{x},\mathbf{v})$.

Special cases of the dynamics~(\ref{NoSymm}) have been derived in the study of decoherence by various authors.  In~\cite{Joos}, the authors discuss the reduced dynamics $\Gamma_{t}$ for a spinless particle interacting with a gas under the assumptions that the reservoir of gas particles is translation invariant, interaction particles from the reservoir are in an ensemble of momentum states (commuting with the momentum operator),  the reservoir is not effected by collisions with the particle, collisions are instantaneous, and an  additional length scale assumption about the collisions the particle receives.     In the three dimensional case, the derived Schr\"odinger dynamics take the form:
\begin{align}\label{Joos}
\frac{d}{dt}\Gamma_{t}(\rho)= i[ |\vec{K}|^{2} ,\Gamma_{t}(\rho)]-\frac{c}{2}\sum_{j=1}^{3}[X_{j} ,[X_{j} ,\Gamma_{t}(\rho)]].
\end{align}
The first term on the right is merely the free dynamics generator, but the second term on the right represents the stochasticity introduced by the reservoir.   Equation~(\ref{Joos}) describes a free particle interrupted by Wiener motion of jumps in momentum.   Notice that the generator has the Lindblad form with irreversible part: $L(\rho)=X\rho X-\frac{1}{2}X^{2}\rho-\frac{1}{2}\rho X^{2}$.  Looking at operator elements in the $x$-basis, $L(\rho)(x_{1},x_{2})= -\frac{1}{2}(x_{1}-x_{2})^{2}\rho(\mathbf{x}_{1},\mathbf{x}_{2})$, so the stochastic term indeed seems to generate an exponential vanishing of off diagonal entries.  Intuitively this  effect, however, is somewhat mitigated by spreading out from the free dynamical term.   An analysis of the decoherence of this model in dimension one is studied in~\cite{Joos} and also in~\cite{Morikawa}, where some additional terms in the Lindblad form corresponding to a harmonic oscillator potential and a friction term are also considered.  In the analysis of~\cite{Joos, Morikawa}, it assumed that the initial density operator $\rho$ has a Gaussian form:
$$ \rho=\frac{2\sqrt{C}}{\sqrt{\pi}}e^{-A(x_{1}-x_{2})^{2}-iB(x_{1}^{2}-x_{2}^{2})-C(x_{1}+x_{2})^{2}-iD(x_{1}-x_{2})-E(x_{1}+x_{2})-F},$$
where all constants $A,\cdots, F$ are real, $A\geq C>0$, and $F=\frac{E^{2}}{4C}$.    $\frac{1}{\sqrt{8C}}$ is the standard deviation of the Gaussian state in the position variable.    The quantity $\frac{1}{\sqrt{8A}}$ is interpreted as the coherence length of the state.   In quantum optics, the coherence length  is the approximate length at which different parts of the wave packet interfere.   The authors in~\cite{Joos} use the fact that the dynamics $\Gamma_{t}$ maps Gaussian density operators to Gaussian density operators and derives differential equations for the coefficients $A_{t},\cdots, F_{t}$.   The relevant quantity for the study of decoherence  is the asymptotics of the ratio $\sqrt{\frac{C_{t}}{A_{t}}}$ (which is equal to the coherence length divided by the standard deviation at time $t$).   For the model~(\ref{Joos}), the asymptotics are $\sqrt{\frac{C_{t}}{A_{t}}}\sim ct^{-2}$ for some constant $c$.    

In~\cite{ESL} there is derivation closely related to that in~\cite{Joos}, but without the short length scale assumption.    The derived  dynamics  $\Gamma_{t}$ satisfy a differential equation which can be written
\begin{align}\label{Gallis}
\frac{d}{dt}\Gamma_{t}(\rho)= i[ |\vec{K}|^{2} ,\Gamma_{t}(\rho)]+\int n(\mathbf{k})d\mathbf{k} (e^{i\mathbf{k}\vec{X}}\Gamma_{t}(\rho)e^{-i\mathbf{k}\vec{X}}-\Gamma_{t}(\rho)   ),
\end{align} 
 where $n(k)$ is a positive density.   These dynamics describe a free particle with a Poisson field of jumps in momentum, where  jumps by $\mathbf{k}$ in occur  with rate $n(\mathbf{k})d\mathbf{k}$.    An equation of this form was originally introduced in~\cite{Ghirardi} as a fundamental alternative to the Schr\"odinger equation rather than an effective reduced dynamics for a particle interacting with an environment.    The dynamics have also have been used to make quantified comparisons with the results of experiments~\cite{Sipe, Alicki}.   For a general discussion of decoherence with an emphasis on these models see~\cite{Decoh}.

The dynamics described by~(\ref{Joos}) and~(\ref{Gallis}) both share the property that they correspond to an environment that is homogenous.  In fact they both satisfy the covariance relation
\begin{align}\label{Galiliean}
\Gamma_{t}(W^{*}_{(x,k)}\rho W_{(x,k)})=W_{(x+tk,k)}^{*}\Gamma_{t}(\rho)W_{(x+tk,k)}, 
\end{align}
 for all Weyl operators $W_{(x,k)}$.   This follows because conjugation by $W_{x,k}$, which corresponds to shift in phase space, commutes with the noise part of the generators.  Moreover, if $F_{t}$ is the free evolution generated by $i[K^{2}, \cdot]$, then $F_{t}(W^{*}_{(x,k)}\rho W_{(x,k)})= W^{*}_{(x+tk,k)}F_{t}(\rho) W_{(x+tk,k)}$.  Hence, even after time evolution, conjugation by Weyl operators commutes with the noise.    A Schr\"odinger dynamics $\Gamma_{t}$ satisfying~(\ref{Galiliean}) is said to be  the Galilean covariant.

    Intuitively, a Galilean covariant semigroup  corresponds to a free particle traveling in a random environment that is invariant with respect to translations in phase space.   In other words,  the probability of the particle undergoing a sudden shift  $(\Delta(\mathbf{x}),\Delta(\mathbf{k}))$ in its position and momentum is independent of its location in phase space.      In~\cite{Holevo}, there is a complete characterization of these processes in terms of their Lindblad form with an additional assumption that the dynamics satisfies rotational covariance $\Gamma_{t}(U_{\sigma}^{*}\rho U_{\sigma})=U_{\sigma}^{*}\Gamma_{t}(\rho)U_{\sigma}$, where $\sigma\in SO_{3}$ and $(U_{\sigma}f)(\mathbf{x})=f(\sigma \mathbf{x})$.    Although Holevo worked in the Heisenberg representation, in the Schr\"odinger representation the dynamics formally satisfy:    
\begin{multline}\label{Holevo}
\frac{d}{dt}\Gamma_{t}(\rho)=  i[ |\vec{K}|^{2} ,\Gamma_{t}(\rho)] -\frac{1}{2}\sum_{i,j=1}^{3}(c^{x,x}[X_{i} ,[X_{j} ,\Gamma_{t}(\rho)]] +c^{x,k}[X_{i} , [K_{j} , \Gamma_{t}(\rho)]] \\ +c^{k,x}[K_{i} , [X_{j} , \Gamma_{t}(\rho)]] +c^{k,k}[K_{i}, [K_{j},\Gamma_{t}(\rho)]] )   +\int d\mu(\mathbf{x},\mathbf{k}) [W_{\mathbf{x,}\mathbf{k}} \Gamma_{t}(\rho)W_{\mathbf{x},\mathbf{k}}^{*}-\Gamma_{t}(\rho) ],
\end{multline}
where the matrix $\begin{pmatrix} c^{x,x} & c^{x.k}\\ c^{k,x} & c^{k,k}\end{pmatrix}$ is real-valued and semi-positive definite, the measure $\mu$ is on $\R^{3}\times \R^{3}-\{0\}$ has the rotational invariance $\mu(\mathbf{x},\mathbf{k})=\mu(\sigma\mathbf{x},\sigma\mathbf{k})$, and the measure has the Levy condition:
\begin{align}\label{MeasCond}
 \int d\mu(\mathbf{x},\mathbf{k})  \frac{|\mathbf{x}|^{2}+|\mathbf{k}|^{2}}{1+|\mathbf{x}|^{2}+|\mathbf{k}|^{2}}<\infty.
 \end{align}
For the integration in~(\ref{Holevo}) to make sense, the integration is taken over spheres centered at the origin first and then in the radial direction (to get a quadratic weight from the integrand near zero).    Looking at the application of Proposition (2) in~\cite{Holevo} to the proof of the Theorem on page 1819 of~\cite{Holevo}, we can see that the analogous results hold when the rotational invariance is removed.    In the case when the rotational symmetry is replaced by just the origin symmetry $d\mu(\mathbf{x},\mathbf{k})=d\mu(-\mathbf{x},-\mathbf{k})$, the corresponding dynamics can be written in the form~(\ref{NoSymm}).  

  The case in which there is only a Poisson term and $\mu(x,k)=\delta(x) \nu(k)$ corresponds to the form derived in~\cite{ESL}, and the case in which there is no Poisson term and only the $c^{x,x}$ quadratic term is non-zero is the model derived in~\cite{Joos}.      Larger classes of dynamics which notably  do not include the energy blow up implied by the form~(\ref{Momentum}) are derived in~\cite{Vacchini0} from scattering arguments.  For a survey of various dynamical semigroups corresponding to spatially invariant environments, see~\cite{Vacchini}.    

One interesting aspect of Galilean covariant dynamics is their constructibility using classical stationary stochastic processes with independent increments.   The dynamics $\Gamma_{t}$ can be constructed as 
\begin{align}\label{Construct}
\Gamma_{t}(\rho)=\mathbb{E}[ W_{(\tilde{x}_{t}+\int_{0}^{t}ds\tilde{k}_{s}, \tilde{k}_{t}) } V_{t}\rho V_{t}^{*} W^{*}_{(\tilde{x}_{t}+\int_{0}^{t}ds\tilde{k}_{s}, \tilde{k}_{t}) }],
\end{align}
 where $V_{t}$ is the unitary group is  generated by $|\vec{K}|^{2}$ (free dynamics) and $( \tilde{x}_{t} , \tilde{k}_{t})  $ is a stationary stochastic Levy process with characteristic functions.  
  $$\varphi_{(\tilde{x}_{t},\tilde{k}_{t})}(\mathbf{q},\mathbf{p})= e^{t \mathit{l}(\mathbf{q},\mathbf{p})},   $$
 where
$$\mathit{l}(\mathbf{q},\mathbf{p})= -\frac{1}{2}c^{x,x}|\mathbf{p}|^{2}-c^{x,k}\mathbf{q}\cdot\mathbf{p}-\frac{1}{2}c^{k,k}|\mathbf{p}|^{2} + \int d\mu(\mathbf{x},\mathbf{k})(e^{i\mathbf{p}\cdot \mathbf{x}+i\mathbf{q}\cdot \mathbf{k}}-1).$$
  Note that we have stated the result~(\ref{Construct}) for the dynamics $\Gamma_{t}$, but the construction in~\cite{Holevo} was made for the adjoint dynamics  $\Gamma_{t}^{*}=\Phi_{t}$ (Heisenberg representation).   The dynamics $\Gamma_{t}$ are thus a statistical average over certain unitary trajectories constructed using the Weyl operators and the free unitary dynamics $U_{t}$.   

The closed factorized form of the characteristic function as found in~(\ref{LevyKhinchin}) for the quantum characteristic function of the covariant dynamics is a consequence of the constructibility of the dynamics $\Gamma_{t}$ using only conjugation by Weyl operators and the $V_{t}$'s.     It is shown in~\cite{Cov} that this implies that Weyl operators evolved under the adjoint dynamics and can be explicitly computed as:
 \begin{align}\label{PhiExplicit}
 \Phi_{t}(W_{(\mathbf{q},\mathbf{p})})= e^{\int_{0}^{t}ds\mathit{l}(\mathbf{q}+(t-s)\mathbf{p},\mathbf{p})  } W_{(\mathbf{q}+t\mathbf{p},\mathbf{p})}.
 \end{align}
A discussion of the dilation of the full collection of processes described by Equation~(\ref{Holevo}) can be found in~\cite{Cov}.   For a larger discussion of dilation of quantum semigroups using classical noise see~\cite{Dissipate}.

In our analysis, we make use of the fact that the map of a trace class operator $\rho$ to its quantum characteristic function $\varphi_{\rho}(\mathbf{q},\mathbf{p})=\Tr[W_{(\mathbf{q},\mathbf{p})}\rho]$ extends to an isometry of  Hilbert-Schmidt operators to  $L^{2}(\R^{d}\times \R^{d}, \frac{1}{(2\pi)^{d}}d\mathbf{q}d\mathbf{p})$.   The dynamics formally satisfying~(\ref{NoSymm}) has a closed expression for the time-evolved quantum characteristic functions $\varphi_{\Gamma_{t}(\rho)}$:
\begin{align} \label{LevyKhinchin}
\varphi_{\Gamma_{t}(\rho)}(\mathbf{q},\mathbf{p})=e^{ \int_{0}^{t}ds \big[ -\frac{1}{2} \left\langle \begin{tiny} \begin{pmatrix} \mathbf{q}+(t-s)\mathbf{p}\\  \mathbf{p} \end{pmatrix}\end{tiny}|\large{A}\begin{tiny}\begin{pmatrix} \mathbf{q}+(t-s)\mathbf{p}\\  \mathbf{p} \end{pmatrix}\end{tiny}\right\rangle
+\psi_{\mu}(\mathbf{q}+(t-s)\mathbf{p},\text{ }\mathbf{p})    \big]}\varphi_{\rho}(\mathbf{q}+t\mathbf{p},\mathbf{p}),
\end{align}
\begin{align}
\text{with }  \psi_{\mu}(\mathbf{q},\mathbf{p})=\int d\mu(\mathbf{x},\mathbf{k}) \big(\textup{cos}(\mathbf{q}\cdot \mathbf{k}+\mathbf{p}\cdot \mathbf{x}) -1\big).
\end{align}
  It is shown that the expression $\psi_{\mu}(\mathbf{q},\mathbf{p})$ can be effectively replaced for the sake of computing the asympototics of $S_{\vec{X}}(\Gamma_{t}(\rho))$  with the quadratic form from the second-order Taylor expansion of $ \psi_{\mu}(\mathbf{q},\mathbf{p})$ at the origin $(\mathbf{q},\mathbf{p})=(0,0)$.   This approximation has the flavor of a central limit theorem where the Poissonian noise can be replaced by a Gaussian noise and can be seen in~(\ref{Relaxation}).    Laplace's method makes it possible to find the asymptotics of quantities needed to calculate $S_{\vec{X}}(\Gamma_{t}(\rho))$.  

Note that the equation~(\ref{LevyKhinchin}) is identical to the characteristic function for the stochastic process $(\tilde{x}_{t}+\int_{0}^{t}ds\tilde{k}_{s},\tilde{k}_{t})$ except that $\varphi_{\rho}(\mathbf{q},\mathbf{p})$ should be replaced by the characteristic function for $(\tilde{x}_{0},\tilde{k}_{0})$ (in the general case, although we have assumed that the process began as $\delta$ distributed at the phase space origin).  The non-classicality arises through the initial characteristic function $\varphi_{\rho}(\mathbf{q},\mathbf{p})$ which will not have the form of a  characteristic function for a classical probability distribution over phase space.    However, as $t\rightarrow \infty$,  the multiplication factor in~(\ref{LevyKhinchin}) vanishes away from origin where $\varphi_{\rho}(\mathbf{q},\mathbf{p})$  takes the generic value of one.   This is the basic reasoning behind the convergence~(\ref{Class}).

This article is organized as follows: Section~\ref{SCI} gives a general discussion of coherence indices of the type~\ref{Commutivity}, Section~\ref{CQDS} gives a brief background on the meaning behind the formal Lindblad equations with unbounded generators as found in the work of Holevo~\cite{Holevo, Dissipate, Cov, Unbounded}, and  Section~\ref{Main} contains the main results of this article.

\section{State Coherence Indices }\label{SCI}
Let $\Hi$ be a complex Hilbert space and $A_{j}$, $j=1,\dots d$ be a family of self-adjoint operators with essential domains $D_{j}$,  and let $\rho$ be a density operator such that $A_{j}\rho$ is trace class ($A_{j}\rho\in \TC_{1}(\Hi)$).   Define $C_{(A_{j})}(\rho)$ and $D_{(A_{j})}(\rho)$ through the following formulas
\begin{align}
C_{(A_{j})}(\rho)=\frac{1}{\|\rho\|_{2}} \Big( \sum_{j=1}^{d}\| [A_{j},\rho]\|_{2}^{2}\Big)^{\frac{1}{2}}
\end{align}
and
\begin{align}
D_{(A_{j})}(\rho)=\frac{1}{\|\rho\|_{2}} \Big(  \sum_{j=1}^{d}\| \{A_{j}-\Tr[A_{j}\rho],\rho\}\|_{2}^{2}\Big)^{\frac{1}{2}}.
\end{align}
The operator $A_{j}\rho$ is defined through the bounded bilinear form $B(g,f)=\langle A_{j}g|\rho f\rangle$.  Notice that $\rho$ maps arbitrary vectors $f$ to the domain of $A_{j}$.

$D_{(A_{j})}(\rho)$ is intended as a sort of  standard deviation for the operators $A_{j}$ in the state $\rho$, while   $C_{(A_{j})}(\rho)$ gives some sort of measure of how close the family of observables $A_{j}$ are to commuting with the state $\rho$.  

\begin{definition}\label{IndexDef}
Let $A_{j}$ be self-adjoint operator with dense domains $D_{j}$ and $\rho \in \TC_{1}(\Hi)$ be a state such that $A_{j}\rho\in \TC(\Hi)$ for each $j$.   If $D_{(A_{j})}(\rho)\neq 0$, then the index $S_{(A_{j})}(\rho)$ of the family $(A_{j})$ with respect to the state $\rho$ is defined as
$$
S_{(A_{j})}(\rho)=\frac{C_{(A_{j})}(\rho)}{ D_{(A_{j})}(\rho)}.
$$
\end{definition}

If the observables $A_{i}$ have some form of units (e.g. length, energy), then the index yields a dimensionless parameter related to the commutativity of the observables $A_{i}$ with respect to the state $\rho$.  The following proposition gives a few basic facts about the index  $S_{(A_{i})}(\rho)$.

\begin{proposition}\label{Index}
Let $\Hi$ be a Hilbert space,  $\rho $ be a density operator, and $(A_{j})$, $j=1,\dots d$ be a family of self-adjoint operator with domains $D_{j}$ such that $A_{j}\rho $ is trace class.  

\begin{enumerate}

\item  \label{Index1}$S_{(A_{j})}(\rho) \in [0,1]$. 

\item  \label{Index2} $S_{(A_{j})}(\rho)=0$ iff $\rho$ commutes with every $A_{j}$.

\item   \label{Index3} If $\rho=|f\rangle\langle f|$ is pure and $f$ is not eigenstate of $A_{j}$ for all $j=1,\cdots d$, then  $S_{(A_{j})}(\rho)=1$.

\end{enumerate}

\end{proposition}

\begin{example}\label{PosExp}
 For  a $1$-dimensional Gaussian state:
$$\rho= \frac{2\sqrt{C}}{\sqrt{\pi}}e^{-A(x_{1}-x_{2})^{2}-iB(x_{1}^{2}-x_{2}^{2})-C(x_{1}+x_{2})^{2}-iD(x_{1}-x_{2})-E(x_{1}+x_{2})-F},$$
$C_{X}(\rho)=\frac{1}{2\sqrt{A} }$, $D_{X}( \rho)=\frac{1}{2\sqrt{C} }$, and the quantity $S_{X}(\rho)=(\frac{C}{A})^{\frac{1}{2}}$,  agrees with  the index used in~\cite{Morikawa, Decoh}.   
\end{example}

Although $C_{X}(\rho)$ and $D_{X}(\rho)$ differes from  the quantities $\frac{1}{\sqrt{8A}}$ and $\frac{1}{\sqrt{8C}}$ interpreted as the coherence length and the standard deviation in~\cite{Morikawa} by a factor of $2^{-\frac{1}{2}}$ for Gaussian density operators, $C_{X}(\rho)$ are $D_{X}(\rho)$ are not amenable to an interpretation of this sort for a general state $\rho$.      The  squaring of an expression involving $\rho$ as found in the formula for $S_{\vec{X}}(\rho)$ can give a skewed weight for the probability weights of events.   The following example gives an extreme situation where this becomes apparent.

\begin{example} 
Let $\Hi=L^{2}(\R)$, $\phi_{m}(x)=(\int_{A_{m}}1dx)^{-\frac{1}{2}}1_{A_{m}} $ where  $A_{m}=[\frac{6}{\pi^{2}}\sum_{r=1}^{m-1}\frac{1}{r^{2}}, \frac{6}{\pi^{2}}\sum_{r=1}^{m}\frac{1}{r^{2}})  $.   Define  the density operators $\rho_{n}= \sum_{m} \lambda_{n,m}|\phi_{m}\rangle \langle \phi_{m}| $, with $\lambda_{n,1}=\frac{1}{n}$, $\lambda_{n,m}=\frac{1}{n^{3}}$ for $2 \leq m \leq n^{3}-n^{2}+1$ and $\lambda_{n,m}= 0$ otherwise.  We can calculate the numerator of $S_{x}(\rho_{n})$ using the following,
\begin{align*}
 \|[X,\rho_{n}]\|_{2}^{2}  = \frac{1}{n^{2}}(\langle \phi_{1}|X^{2} \phi_{1} \rangle - \langle \phi_{1}|X \phi_{1}\rangle^{2})+\mathit{O}(\frac{1}{n^{3}}).
 \end{align*}
 Moreover we can approximate the denominator of $S_{x}(\rho_{n})$ as,
  \begin{align*}
 \|\{X,\rho_{n}\}\|_{2}^{2}  =  \frac{1}{n^{2}}\langle \phi_{1}|(X-1)^{2} \phi_{1} \rangle+\frac{1}{n^{2}}\langle \phi_{1}|(X-1)\phi_{1}\rangle^{2}+\mathit{O}(\frac{1}{n^{3}}) ,
\end{align*}
  Hence,
\begin{align*}
S_{x}(\rho_{n}) \sim \frac{(\langle \phi_{1}|X^{2} \phi_{1} \rangle - \langle \phi_{1}|X \phi_{1}\rangle^{2})^{\frac{1}{2}} }{ (\langle \phi_{1}|(X-1)^{2} \phi_{1} \rangle+\langle \phi_{1}|(X-1)\phi_{1}\rangle^{2})^{\frac{1}{2}} }
\end{align*}
The above expression depends only on the first state $\phi_{1}$ even though this state has a weight of only $\frac{1}{n^{2}}$, which is a diminishing fraction of the total weight.    

\end{example} 
In general, just as for classical diffusion processes, only states of very specific forms can occur  after they have been acted upon by an irreversible environment  $\rho\rightarrow \Gamma_{t}(\rho)$.   The states that are likely to occur depend on the nature of the environment.    Our analysis, in Section~\ref{Main} essentially relies on the fact that when stochastic shifts in momentum are present, then after sufficient time $\Gamma_{t}(\rho)$ becomes essentially Gaussian--which is the statement of Theorem~(\ref{Relaxation}).     Thus for those dynamics, $S_{\vec{X}}(\rho)$ is asymptotically expected to serve well as a coherence index.   In the case where there are only stochastic shifts in position, the state $\Gamma_{t}(\rho)$ becomes in some sense only partially Gaussian since the asymptotic characteristic function~(\ref{LevyKhinchin}) will only be forced to be quadratic in the exponent with respect to the $\vec{p}$ variables.   This seems apparent in the asymptotics~(\ref{PosOnly}), since the constant
$$\frac{ (  \int d\mathbf{k}|\rho(\mathbf{k},\mathbf{k})|^{2} | \mathbf{k}-E[\vec{K}\rho]|^{2} )^{\frac{1}{2}}}{(\int d\mathbf{k}|\rho(\mathbf{k},\mathbf{k})|^{2})^{\frac{1}{2}} } $$
has the strange squaring of $ \rho(\mathbf{k},\mathbf{k})$.    With a more accurate formula for the coherence length divided by the standard deviation in position, we expect that this constant would be replaced by a variance formula for the probability density $\rho(\mathbf{k},\mathbf{k})$.    In general, for other types dynamics we expect $S_{\vec{X}}(\rho_{t})$ to capture the correct power law even if the constants are not completely natural.  

In the proposition below, we give useful expressions for the quantities need to compute $S_{\vec{X}}(\rho)$ and $S_{\vec{K}}(\rho)$.

\begin{proposition}\label{CharSCI}
Let $\rho$ be a state such that $ J\rho \in \TC_{1}(L^{2}(\R^{d}))$ for any \\ $J\in\{X_{1}, \cdots, X_{d}, K_{1},\cdots, K_{d}\}$, and define 
$$\mathbf{v}_{\mathbf{p}}=(\nabla_{\mathbf{p}}\varphi_{\rho})(0,0) \text{ and } \mathbf{v}_{\mathbf{q}}=(\nabla_{\mathbf{q}}\varphi_{\rho})(0,0).$$
Then for the vector of position observables $\vec{X}$,
$$\|[\vec{X},\rho]\|_{2}= \big( \int d\mathbf{q}d\mathbf{p} |\mathbf{q}|^{2}|\varphi_{\rho}(\mathbf{q},\mathbf{p})|^{2} \big)^{\frac{1}{2}}, \text{and } \|\{\vec{X},\rho\}\|_{2}=\big( \int d\mathbf{q}d\mathbf{p} |(\nabla_{\mathbf{p}}-\mathbf{v}_{\mathbf{p}}) \varphi_{\rho}(\mathbf{q},\mathbf{p})|^{2}\big)^{\frac{1}{2}}.  $$
For the momentum variable $\vec{K}$,
$$\|[\vec{K},\rho]\|_{2}= ( \int d\mathbf{q}d\mathbf{p} |\mathbf{p}|^{2}|\varphi_{\rho}(\mathbf{q},\mathbf{p})|^{2}\big)^{\frac{1}{2}}, \text{and } \|\{\vec{K},\rho\}\|_{2}=\big( \int d\mathbf{q}d\mathbf{p} |(\nabla_{\mathbf{q}}-\mathbf{v}_{\mathbf{q}} )\varphi_{\rho}(\mathbf{q},\mathbf{p}) |^{2}  \big)^{\frac{1}{2}}   $$
\end{proposition}

\begin{proof}
By~(\ref{Isometry}), quantum characteristic functions define an isometry from Hilbert-Schmidt class operators to functions in $L^{2}(\R^{d}\times \R^{d}, \frac{1}{(2\pi)^{d}}d\mathbf{p}d\mathbf{q})$ (Lebesgue measure on phase space multiplied by a factor $\frac{1}{(2\pi)^{d}}$).   Hence we have that 
$$ \Tr[ -[\vec{X},\rho]^{2}] = \frac{1}{(2\pi)^{d}}\int d\mathbf{q}d\mathbf{p} \sum_{j}|\varphi_{i[X_{j},\rho]}(\mathbf{q},\mathbf{p})|^{2}    $$
By definition $\varphi_{i[\vec{X},\rho]}(\mathbf{q},\mathbf{p})=\Tr[ e^{i(\mathbf{q}\cdot \vec{K}+\mathbf{p}\cdot \vec{X})}i[\vec{X},\rho]]$.  However, we can write $i[\vec{X},\rho]=\nabla_{\vec{a}}|_{\vec{a}=0}W_{\vec{a}}^{*}\rho W_{\vec{a}}$ where convergence for the limits $   \frac{1}{h}(W_{he_{i}}^{*}\rho W_{he_{i}}-\rho)\rightarrow i[X_{i},\rho]$ takes place in the trace norm by Lemma~(\ref{Use}).    Since the convergence is in the trace norm it follows that we can commute the limit with the trace in the following computation:
\begin{multline*}
\Tr[ e^{i(\mathbf{q}\cdot \vec{K}+\mathbf{p}\cdot \vec{X})}i[\vec{X},\rho]]=\nabla_{\vec{a}}|_{\vec{a}=0}\Tr[e^{i(\mathbf{q}\cdot \vec{K}+\mathbf{p}\cdot \vec{X})}W_{(\vec{a},0)}^{*}\rho W_{(\vec{a},0)}] \\ = \nabla_{(\vec{a},0)}|_{\vec{a}=0}\Tr[ W_{(\vec{a},0)}e^{i(\mathbf{q}\cdot \vec{K}+\mathbf{p}\cdot \vec{X})}W_{(\vec{a},0)}^{*}\rho ]= \nabla_{\vec{a}}|_{\vec{a}=0}\Tr[ e^{i( (\mathbf{q}\cdot (\vec{K}-\vec{a})+\mathbf{p}\cdot \vec{X})}\rho ] \\ = \nabla_{\vec{a}}|_{\vec{a}=0}(e^{i\mathbf{q}\cdot \vec{a}})\Tr[ e^{i( (\mathbf{q}\cdot \vec{K}+\mathbf{p}\cdot \vec{X})}\rho ]= -i\mathbf{q}\varphi_{\rho}(\mathbf{q},\mathbf{p})
\end{multline*}
   Hence we can conclude that
$$\|[\vec{X},\rho]\|_{2}=\Big( \int d\mathbf{q}d\mathbf{p} |\mathbf{q}\varphi_{\rho}(\mathbf{q},\mathbf{p})|^{2}\Big)^{\frac{1}{2}}. $$
The quatities $\|\{\vec{X},\rho\}\|_{2}$, $\|[\vec{K},\rho]\|_{2}$, and $\|\{\vec{K},\rho\}\|_{2}$ have  similar arguments.

\end{proof}

\section{Covariant Quantum Dynamical Semigroups }\label{CQDS}
For the purposes of the decoherence analysis in the next section, we work with the characteristic functions~(\ref{LevyKhinchin}), using~(\ref{CharSCI}).  In this section, we discuss the meaning behind the formal Markovian master equations with unbounded generators discussed in the introduction.       For a more in depth view of this topic see~\cite{Holevo,Dissipate,Cov,Unbounded} and further references.    We finish up by making a few comments on the action of the dynamics from the perspective of characteristic functions. 
 
Given an Hilbert space $\Hi$, a dynamics can be seen as a collection of completely positive maps (cpm's) in the Schr\"odinger picture  acting on trace class operators $\Gamma_{t}: \TC_{1}(\Hi)\rightarrow \TC_{1}(\Hi)$, or in the Heisenberg picture acting on bounded operators $\Phi_{t}:\Bi(\Hi)\rightarrow \Bi(\Hi)$.    The dynamics $\Gamma_{t}$ and $\Phi_{t}$ are related through the trace formula:
 \begin{align}\label{Adjoint}
 \Tr[ \Gamma_{t}(\rho)G]=\Tr[\rho\Phi_{t}(G)].
 \end{align}
Since $\TC(\Hi)^{*}=\Bi(\Hi)$, the maps $\Gamma_{t}$ are pre-adjoint to $\Phi_{t}$.   Although physicists working in quantum optics tend to work in the Schr\"odinger picture,  those working on existence and uniqueness  of Lindblad type equations  tend to use the adjoint dynamics.  Through Equation~(\ref{Adjoint}) either dynamics can be constructed using the other.   Also, the dynamics $\Gamma_{t}$  is trace preserving iff the adjoint dynamcis $\Phi_{t}$ is unital (i.e. $\Phi_{t}(I)=I$ for all $t$).   The maps $\Phi_{t}$ are said to form a \text{dynamical semigroup} if $\Phi_{t}\Phi_{s}=\Phi_{t+s}$, and $\Tr[\rho\Phi_{t}(G)]$ is continuous (i.e. $\text{weak}^{*}$-continuous).

In~\cite{Cov}, Holevo studies a dynamics $\Phi_{t}$ operating on $\Bi(L^{2}(\R^{3}))$ in the Heisenberg representation and formally satisfying:
\begin{multline}\label{Holevo2}
\frac{d}{dt}\Phi_{t}(G)=  -i[ |\vec{K}|^{2} ,\Phi_{t}(G)] -\frac{1}{2}\sum_{j}(c^{x,x}[X_{j} ,[X_{j} ,\Phi_{t}(G)]] -c^{x,k}[X_{j} , [K_{j} , \Phi_{t}(G)]] \\-c^{k,x}[K_{j} , [X_{j} , \Phi_{t}(G)]] -c^{k,k}[K, [K,\Phi_{t}(G)]] )  +\int d\mu(\mathbf{x},\mathbf{k}) [W_{\mathbf{x},\mathbf{k}}\Phi_{t}(G)W_{\mathbf{x},\mathbf{k}}^{*}-\Phi_{t}(G)],
\end{multline}
 where $\begin{pmatrix}c^{x,x} & c^{x,k}\\ c^{k,x} & c^{k,k} \end{pmatrix}$ is a positive matrix with real valued entries and  $\mu$ is a measure on $\R^{3}\times \R^{3}$ satisfying the L\'evy condition $\int d\mu(\mathbf{x},\mathbf{k})\frac{|\mathbf{x}|^{2}+|\mathbf{k}|^{2}}{1+|\mathbf{x}|^{2}+|\mathbf{k}|^{2}}<\infty$ and the rotational invariance $\mu(\mathbf{x},\mathbf{k})=\mu(\sigma\mathbf{x},\sigma\mathbf{k})$ for $\sigma\in SO_{3}$.

Since the Lindblad Equation~(\ref{Holevo2}) has an unbounded generator, the classic result~\cite{Lindblad} guaranteeing the existence and uniqueness of a norm continuous adjoint semigroup $\Phi_{t}$ of completely positive maps satisfying  $\Phi_{t}(I)=I$ does not apply.   Just as in the case of generators of unitary groups, unbounded generators of Markovian semi-groups require extra care to define and pose new technical difficulties.   One approach for dealing with these technical issues  is the introduction of a \textit{form generator}.

\begin{definition} 
Let $D\subset \Hi$ be dense.   A form generator is a linear map $L:D\times \Bi(\Hi)\times D\rightarrow \C$ such that for $f,g\in D$ and $G\in \Bi(\Hi)$,
\begin{enumerate}
\item
$$\mathcal{L}(g; G; f)= \overline{ \mathcal{L}(f; G^{*}; g) }  $$

\item
$$\sum_{l,j}\mathcal{L}(f_{l}; G^{*}_{l}G_{j}; f_{j})\geq 0 \text{ when } \sum_{j} G_{j}f_{j}=0 $$

\item
For any fixed $g,f$, $\mathcal{L}(g; G; f)$ is continuous in $G$ with respect to the strong topology over any bounded subset of $\Bi(\Hi)$.

\end{enumerate}

\end{definition}

A form generator $\mathcal{L}$ is said to be \textit{unital} if $\mathcal{L}(g;I;g)=0$ for all $f,g\in D$.    The definition for the form generators is inspired by the form of a bounded Lindblad generator.  In~\cite{Holevo}, it is shown that for any form generator $\mathcal{L}$ there exist operators $L_{j}$, $j\in \N$ and $B$ with domains including $D$ such that
 \begin{align*}
 \mathcal{L}(g;G; f)=\sum_{j}\langle L_{j} g| G L_{j} f\rangle-\frac{1}{2}\langle B g | G f\rangle -\frac{1}{2}\langle g| G B f\rangle 
 \end{align*}

Given a form generator $\mathcal{L}$, we can then ask if there is a process $\Phi_{t}$ satisfying $\Phi_{0}(G)=G$ and  
\begin{align}
\frac{d}{dt}\langle g| \Phi_{t}(G) f \rangle=\mathcal{L}(g; G ; f),
\end{align}
where $g,f\in D$ and $G\in \Bi(\Hi)$ and some regularity properties are assumed for $\Gamma_{t}(G)$.    An important criterion used for the construction of solutions to this equation is that $B$ is a maximal accretive operator.   By an analogous result to Stone's Theorem~\cite{Reed}, maximal accretive operators are the generators of strongly continuous semigroups of contractive maps~\cite{Kato}.  In~\cite{Holevo} it is shown that for any unital form generator $\mathcal{L}$ admitting a Lindblad form where the operator $B$ is maximal accretive then there exists a unique \textit{minimal}  dynamical semigroup $\Phi$ to the equation  
\begin{align}\label{Minimal}
\frac{d}{dt}\langle g| \Phi_{t}(G) f\rangle= \mathcal{L}(g; \Phi_{t}(G); f) 
\end{align}
where $\Phi_{0}(G)=G$.   A solution $\Phi_{t}$ to the above equation is said to be \textit{minimal}, if for any other solution $\Phi_{t}^{\prime}$:
$$\Phi_{t}^{\prime}(G)\geq \Phi_{t}(G) \text{ when } G\geq 0.   $$
 Surprisingly, the conservativity of the minimal solution ($\Phi_{t}(I)=I$) is not guaranteed if the form generator in unital.    A general set of necessary and sufficient conditions for guaranteeing conservativity is unknown, and in the literature stringent conditions are assumed in order to prove the conservativity for a specific class of form generators~\cite{Cheb,Holevo}.

For  $f,g\in D=\cap_{\vec{q},\vec{p}}\textup{Dom}(\vec{p}\cdot \vec{K}+\vec{q}\cdot \vec{X})$, the form generator $\mathcal{L}(g; G ; f)$ of the adjoint dynamics corresponding to the formal Equation~(\ref{Holevo2}) has the form:
\begin{align}
\mathcal{L}(g; G ; f)= T_{1}(g;G;f)+T_{2}(g;G;f)+T_{3}(g;G;f),
\end{align}
where
\begin{align*}
T_{1}(g;G;f)= -i\langle  K^{2}g| G f \rangle+i\langle g| G K^{2} f\rangle,
\end{align*}
\begin{multline*}
T_{2}(g;G;f)= \sum_{j=1}^{3}\big( c^{x,x}\langle X_{j} g| G X_{j}f\rangle+c^{x,k}\langle X_{j} g| G K_{j} f \rangle+c^{k,x}\langle K_{j} g| G X_{j} f\rangle \\ +c^{k,k}\langle K_{j} g| G K_{j} f\rangle -\frac{1}{2}\langle (c^{x,x}X_{j}^{2}+c^{k,x}K_{j}X_{j}+c^{x,k}X_{j}K_{j}+c^{k,k}K_{j}^{2})g| G f\rangle \\ -\frac{1}{2}\langle g |G(c^{x,x}X_{j}^{2}+c^{k,x}K_{j}X_{j}+c^{x,k}X_{j}K_{j}+c^{k,k}K_{j}^{2})f\rangle \big),
\end{multline*}
\begin{align*}
T_{3}(g;G;f)=\int d\mu(\mathbf{x},\mathbf{k})\big( \langle W_{\mathbf{x},\mathbf{k}}^{*}g| G W_{\mathbf{x},\mathbf{k}}^{*}f\rangle-\langle g| G f\rangle   \big),
\end{align*}
and the integral is taken over surfaces of equal radius to make the integration well defined.  

In~\cite{Cov}, it is shown that for an $G\in \Bi(\Hi)$ there is a unique conservative dynamical semigroup $\Phi_{t}(G)$ with $\Phi_{0}(G)=G$ and satisfying the equation 
$$\frac{d}{dt}\langle g|\Phi_{t}(G) f\rangle= \mathcal{L}(g; \Phi_{t}(G) f\rangle,$$
 and the dynamics $\Phi_{t}$ have the covariance relations:
$$\Phi_{t}(W_{(\mathbf{q},\mathbf{p})}^{*}G W_{(\mathbf{q},\mathbf{p})})=W_{(\mathbf{q}+t\mathbf{p},\mathbf{p})}^{*}\Phi_{t}(G) W_{(\mathbf{q}+t\mathbf{p},\mathbf{p})}, \text{ and } \Phi_{t}(R_{\sigma}^{*}G R_{\sigma})=R_{\sigma}^{*}\Phi_{t}(G) R_{\sigma}.$$
Conversely, it is shown that any conservative dynamical semigroup satisfying the covariance relations above is the unique  solution to an equation of the form~(\ref{Minimal}).   

Since the dynamics  $\Phi_{t}$ acting on any Weyl operator $W_{\mathbf{q},\mathbf{p}}$ is explicitly computable~(\ref{PhiExplicit}), this implies that the quantum characteristic functions of the predual process $\Gamma_{t}$ are explicitly computable, since  
\begin{multline*}
\varphi_{\Gamma_{t}(\rho)}= \Tr[W_{\mathbf{q},\mathbf{p}}\Gamma_{t}(\rho)]=\Tr[\Phi_{t}(W_{\mathbf{q},\mathbf{p}})\rho]\\ =e^{ \int_{0}^{t} l(q+(t-s)p,p)ds   }\Tr[W_{\mathbf{q}+t\mathbf{p},\mathbf{p}} \rho]= e^{ \int_{0}^{t} l(q+(t-s)p,p)ds   }\varphi_{\rho}(\mathbf{q}+t\mathbf{p},\mathbf{p}). 
\end{multline*}
However, for the free dynamics $F_{t}$ generated by $i[|\vec{K}|^{2},\cdot]$, $\varphi_{F_{t}(\rho)}(\mathbf{q},\mathbf{p})=\varphi_{\rho}(\mathbf{q}+t\mathbf{p},\mathbf{p})$, hence in the formula above we have factorization of the quantum characteristic function with a noise part and a deterministic part.   The  stochastic factor $e^{ \int_{0}^{t}l(q+(t-s)p,p)ds   }$ is a consequence of an analogous construction to~(\ref{Construct}) for the adjoint dynamics $\Phi_{t}$ and basic computations with Weyl operators.

It is useful to think about how the dynamics act in terms of their quantum characteristic functions. 
We can define the action of the dynamics $\mathit{\Gamma}_{t}$ and $\mathcal{F}_{t}$ acting on characteristic functions through the formula:
$$\mathit{\Gamma}_{t}\varphi_{\rho}=\varphi_{\Gamma_{t}(\rho)}, \text{ and } \mathcal{F}_{t}\varphi_{\rho}=\varphi_{F_{t}\rho}.$$
Notice that $\mathit{\Gamma}_{t}$ forms a semigroup of contractive maps on $L^{2}(\R^{d}\times \R^{d})$.  This can be seen through the formula~(\ref{LevyKhinchin}), but follows from more  general considerations.  The quantum characteristic functions define an isometry from $\TC_{2}(L^{2}(\R^{d}))$ to $L^{2}(\R^{d}\times \R^{d}, (2\pi)^{-d}d\mathbf{x}d\mathbf{k})$.   The maps $\Gamma_{t}$ are completely positive and  extend to unital maps (since they satisfy the same equation as the adjoint maps with $i$ replaced by $-i$ in the kinetic part of the Lindblad generator), so 
$$\Gamma_{t}(\rho^{*})\Gamma_{t}(\rho)\leq \Gamma_{t}(\rho^{*}\rho).$$
Taking the trace of both sides and using the isometry
$$\|\mathit{\Gamma}_{t}\varphi_{\rho}\|_{2}\leq \|\varphi_{\rho}\|_{2}.$$

In many cases, we will find it convenient to write
\begin{align}\label{LevyKhinchin2} 
\varphi_{\Gamma_{t}(\rho)}(\mathbf{q},\mathbf{p})=\mathit{F}_{t} \mathit{\Gamma}_{t}^{\prime}\varphi_{\rho}(\mathbf{q},\mathbf{p}),
\end{align}
where $\mathit{\Gamma}_{t}^{\prime}$ is the multiplication operator of the form
\begin{align}\label{LevyKhinchin2p}
\mathit{\Gamma}_{t}^{\prime}= e^{ \int_{0}^{t}ds \big[ -\frac{1}{2} \left\langle \begin{tiny}\begin{pmatrix} \mathbf{q}-s\mathbf{p}\\ \mathbf{p}\end{pmatrix}\end{tiny} |A  \begin{tiny}\begin{pmatrix} \mathbf{q}-s\mathbf{p}\\ \mathbf{p}\end{pmatrix}\end{tiny}\right\rangle  + \psi_{\mu}(\mathbf{q}-s\mathbf{p},\mathbf{p}) \big]}. 
\end{align}

\section{Decoherence Rates for Covariant Dynamics }\label{Main}
In the section, we will compute decoherence rates for cases of covariant dynamics: where there is only stochastic shifts in momentum, only stochastic shift is position, and an active presence of both stochastic shift in momentum and position.   The analysis is not  an exaustive case analysis, since we always make an assumption such that either $A^{x,x}, A^{k,k}, A$ is positive (rather than just positive semidefinite) or that the measures $\mu$ or $\nu$ have densities.  However, for the situation where $\mu$ is assumed to have second moments, the main situations are covered.    The proof of~(\ref{ThmPoisson}) is essentially what is needed for~(\ref{Relaxation}). 

By the characteristic function isometry, Equation~(\ref{LevyKhinchin2p}), and the fact that $\mathit{F}_{t}$ acts as an isometry on $L^{2}(\R^{d},\R^{d})$, we have that
\begin{align}\label{DynSub2}
\|[\vec{X},\Gamma_{t}(\rho)]\|_{2}= \frac{1}{(2\pi)^{\frac{d}{2}} }\|(\mathbf{q}-t\mathbf{p})\mathit{\Gamma}_{t}^{\prime}(\varphi_{\rho})\|_{2},  \text{ and}
\end{align}
\begin{align}\label{Notmuch}
\|\{\vec{X}-\Tr[\vec{X}\rho], \Gamma_{t}(\rho)\}\|_{2}  = \frac{1}{(2\pi)^{\frac{d}{2}} }\|(t\nabla_{\mathbf{q}}+\nabla_{\mathbf{p}}-\nabla_{\mathbf{p}}\varphi_{\mathit{\Gamma}_{t}(\rho)}(0,0) )\mathit{\Gamma}_{t}^{\prime}(\varphi_{\rho})\|_{2} .
\end{align}
Moreover, by the origin symmetry of the noise, the noise does not change the expectation of the momentum and the position operators from the initial state.   Hence with $\mathit{\Gamma}_{t}=\mathit{F}_{t}\mathit{\Gamma}_{t}^{\prime}$, $E[\vec{X}\rho]=\nabla_{p}\varphi_{\rho}(0,0)$ and $E[\vec{K}\rho]=\nabla_{q}\varphi_{\rho}(0,0)$,
\begin{align}
\nabla_{\mathbf{p}}\varphi_{\mathit{F}_{t}\mathit{\Gamma}_{t}^{\prime}(\rho)}(0,0)=t\nabla_{\mathbf{q}}\varphi_{\mathit{\Gamma}_{t}\rho}(0,0)+\nabla_{p}\varphi_{\mathit{\Gamma}_{t}(\rho)}(0,0)= t\nabla_{\mathbf{q}}\varphi_{\rho}(0,0)+\nabla_{\mathit{p}}\varphi_{ \rho}(0,0).
\end{align}
The last term from Equation~(\ref{Notmuch}) is  bounded from above and below by,  
\begin{multline}\label{DynSub3}
 \frac{1}{(2\pi)^{\frac{d}{2}} }\| [t\nabla_{\mathbf{q}}+\nabla_{\mathbf{p}}, \mathit{\Gamma}_{t}^{\prime}]\varphi_{\rho}\|_{2} \pm \Big( \frac{1}{(2\pi)^{\frac{d}{2}} }\|(t\nabla_{\mathbf{q}}\varphi_{\rho}(0,0)+\nabla_{\mathit{p}}\varphi_{ \rho}(0,0))\mathit{\Gamma}_{t}^{\prime}(\varphi_{\rho})\|_{2} \\ +\frac{1}{(2\pi)^{\frac{d}{2}} }\|\mathit{\Gamma}_{t}(t\nabla_{\mathbf{q}}+\nabla_{\mathbf{p}})\varphi_{\rho}\|_{2} \Big)
\end{multline}
Where for some cases the later term will be seen to be of smaller order.

\begin{proposition}\label{ThmBrown}
Let $\rho$ be a density operator such that $J\rho \in \TC_{1}(L^{2}(\R^{d}))$ for all \\$J\in \{X_{1},\cdots, X_{d},K_{1},\cdots, K_{d}\}$.  In the case when $\varphi_{\Gamma_{t}}(\rho)$ satisfies Equation~(\ref{LevyKhinchin})  with $\mu=0$,
\begin{enumerate}
\item $A^{k,x}=A^{x,k}=A^{k,k}=0$, and  $A^{xx}$ positive, then 
$$S_{\vec{X}}(\Gamma_{t}(\rho))\sim t^{-2}\sqrt{3}\frac{\Tr[(A^{x,x})^{-1}]^{\frac{1}{2}}}{\Tr[A^{x,x}]^{\frac{1}{2}}},  $$
\item $A^{x,x}=A^{k,x}=A^{x,k}=0$, and $A^{k,k}$ positive, then 
$$S_{\vec{X}}(\Gamma_{t}(\rho))\sim t^{-\frac{1}{2} }2^{\frac{1}{2}} \Tr[(A^{k,k})^{-1}]^{\frac{1}{2}} \frac{ (  \int d\mathbf{k}|\rho(\mathbf{k},\mathbf{k})|^{2} | \mathbf{k}-E[\vec{K}\rho]|^{2} )^{\frac{1}{2}}}{(\int d\mathbf{k}|\rho(\mathbf{k},\mathbf{k})|^{2})^{\frac{1}{2}} }, $$
\item  and $A$ is positive, then 
$$S_{\vec{X}}(\Gamma_{t}(\rho))\sim t^{-2}\sqrt{3}\frac{\Tr[(A^{x,x})^{-1}]^{\frac{1}{2}}}{\Tr[A^{x,x}]^{\frac{1}{2}}}.   $$
\end{enumerate}

\end{proposition}

\begin{proof}

For  the numerator of $S_{\vec{X}}(\Gamma_{t}(\rho))$, we can use Equation~(\ref{DynSub2})
$$\|[\vec{X},\Gamma_{t}(\rho)]\|_{2}=\frac{1}{(2\pi)^{\frac{d}{2}} }\|(\mathbf{q}-t\mathbf{p})\mathit{\Gamma}_{t}^{\prime}(\varphi_{\rho})\|_{2},
\text{ where } $$
$$ \mathit{\Gamma}_{t}^{\prime}(\varphi_{\rho})= \frac{1}{(2\pi)^{d} }\int d\mathbf{q}d\mathbf{p}e^{-\int_{0}^{t}ds  \left\langle \begin{tiny} \begin{pmatrix} \mathbf{q}-s\mathbf{p}\\ \mathbf{p} \end{pmatrix} \end{tiny} | A \begin{tiny} \begin{pmatrix} \mathbf{q}-s\mathbf{p}\\ \mathbf{p} \end{pmatrix} \end{tiny} \right\rangle  ds } |\varphi_{\rho}(\mathbf{q},\mathbf{p})|^{2}.$$
Computing the integral in the exponent and rearranging, 
\begin{align*}
\int_{0}^{t}ds \left\langle \begin{tiny} \begin{pmatrix} \mathbf{q}-s\mathbf{p}\\ \mathbf{p} \end{pmatrix} \end{tiny} | A \begin{tiny} \begin{pmatrix} \mathbf{q}-s\mathbf{p}\\ \mathbf{p} \end{pmatrix} \end{tiny} \right\rangle  ds =\frac{t}{4}\left\langle \begin{tiny} \begin{pmatrix} \mathbf{q}\\ \mathbf{p} \end{pmatrix}\end{tiny} | A \begin{tiny} \begin{pmatrix} \mathbf{q}\\ \mathbf{p} \end{pmatrix}\end{tiny} \right\rangle
  +\frac{t^{3}}{3}\left\langle \begin{tiny} \begin{pmatrix}\mathbf{p}-\frac{3}{2t}\mathbf{q} \\ \frac{3}{2t}\mathbf{p} \end{pmatrix} \end{tiny} |A \begin{tiny} \begin{pmatrix}\mathbf{p}-\frac{3}{2t}\mathbf{q} \\ \frac{3}{2t}\mathbf{p} \end{pmatrix} \end{tiny} \right\rangle. 
\end{align*}

Now we will begin case analysis.\\
\noindent \textbf{Case 1}: Since $A^{x,x}$ is positive definite, there exists a unitary $U$ and a diagonal $D$ such that $A^{x,x}=U^{*}DU$.  By changing variables $U\mathbf{q}\rightarrow \mathbf{q}$ and $U(\mathbf{p}-\frac{3}{2t}\mathbf{q})\rightarrow \mathbf{p}$, we can then write
$$\|[\vec{X},\Gamma_{t}(\rho)]\|_{2}^{2}=\frac{1}{(2\pi)^{d}}\int d\mathbf{q}d\mathbf{p}(\frac{1}{4}|\mathbf{q}|^{2}+t^{2}|\mathbf{p}|^{2})e^{-\frac{t}{2}\langle \mathbf{q}| D\mathbf{q}\rangle -\frac{2t^{3}}{3}\langle \mathbf{p}| D\mathbf{p}\rangle  }. $$
By Lemma~(\ref{DerivChar}), $\varphi_{\rho}$ is uniformly continuous with $\varphi_{\rho}(0,0)=1$.  Hence if $\lambda_{i}$ are the entries of $D$ we can apply Laplace's method to calculate the asymptotics of the above expression as
$$   [ \frac{1}{4t}(\frac{1}{\lambda_{1}}+\cdots \frac{1}{\lambda_{d}})+t^{2}\frac{3}{4t^{3}}(\frac{1}{\lambda_{1}}+\cdots \frac{1}{\lambda_{d}}) ] \|\mathit{\Gamma}_{t}(\varphi_{\rho})\|_{2}^{2}, \text{ where } \|\mathit{\Gamma}_{t}(\varphi_{\rho})\|_{2}^{2}\sim  t^{-2d}(\lambda_{1}\cdots \lambda_{d})^{-1}( \frac{\sqrt{3}}{2})^{d}.$$
Hence $\|[\vec{X},\Gamma_{t}(\rho)]\|_{2}\sim t^{-\frac{1}{2} }\Tr[A^{-1}]^{\frac{1}{2}}\|\Gamma_{t}(\rho)\|_{2}$.

To get a hold of the denominator we study the expression $ \frac{1}{\pi^{\frac{d}{2}} }\| [\nabla_{\mathbf{p}}-t\nabla_{\mathbf{q}}, \mathit{\Gamma}_{t}^{\prime}]\varphi_{\rho}\|_{2}$ from~(\ref{DynSub3}) since the other terms have smaller order.  The commutation is between derivatives and a multiplication operator and hence can be explicitly computed:
$$\frac{1}{(2\pi)^{d}}\int d\mathbf{q}d\mathbf{p}|-t^{2}A\mathbf{q}+\frac{t^{3}}{3}A\mathbf{p}|^{2}e^{-\frac{t}{2}A^{x,x}\mathbf{q}^{2}-\frac{2t^{3}}{3}A^{x,x}(\mathbf{p}-\frac{3}{2t}\mathbf{q})^{2} }. $$
If we rewrite $t^{2}\mathbf{q}-\frac{t^{3}}{3}\mathbf{p}=\frac{t^{2}}{2}\mathbf{q} -\frac{t^{3}}{3}(\mathbf{p}-\frac{3}{2t}\mathbf{q})$, then making the same change of variables as above, Laplace's method gives
$$ [ \sum_{j} (\lambda_{j}^{2}(\frac{t^{2}}{2})^{2}\frac{1}{t\lambda_{j}}+\lambda_{j}^{2}(\frac{t^{3}}{3})^{2}\frac{3}{4t^{3}\lambda_{j}}  ] \|\Gamma_{t}(\rho)\|_{2}^{2}.$$

 The other terms from~(\ref{DynSub3}), $\frac{1}{(2\pi)^{\frac{d}{2}} }\|\big(t\nabla_{\mathbf{q}}\varphi_{\rho}(0,0)+\nabla_{\mathbf{p}}\varphi_{\rho}(0,0) \big)\mathit{\Gamma}_{t}^{\prime}(\varphi_{\rho})\|_{2}$  and  $\frac{1}{\pi^{\frac{d}{2}} }\|\mathit{\Gamma}_{t}^{\prime}(t\nabla_{\mathbf{q}}+\nabla_{\mathbf{p}})\varphi_{\rho}(\mathbf{q},\mathbf{p})\|_{2}$, can be at most of order $t$.  For the second term we use that the derivatives of $\varphi_{\rho}$ are continuous and uniformly bounded by Lemma~(\ref{DerivChar}) in order to apply Laplace's method.  Hence  $S_{\vec{X}}(\Gamma_{t}(\rho))\sim t^{-2}\sqrt{3}\frac{\Tr[(A^{x,x})^{-1}]^{\frac{1}{2}}}{\Tr[A^{x,x}]^{\frac{1}{2}}} $. 

\noindent \textbf{Case 2}:\\
Let $A^{k,k}=U^{*}DU$ where $D$ is diagonal with entries $\lambda_{j}$.  By changing variables $U\mathbf{p}\rightarrow \mathbf{p}$, we obtain the expression
$$\frac{1}{(2\pi)^{d} }\int d\mathbf{q}d\mathbf{p}(|\mathbf{q}|^{2}-2t\mathbf{q}\mathbf{p}+t^{2}|\mathbf{p}|^{2} )e^{-t\langle \mathbf{p}|D\mathbf{p}\rangle }|\varphi_{\rho}(\mathbf{q},U^{*}\mathbf{p})|^{2}.$$
However, the third term dominates since $|\mathbf{p}|\sim \frac{1}{\sqrt{t}}$.  
In the limit, $t\rightarrow \infty$, the exponential factor places a weight on the surface $\mathbf{p}=0$.  By 
Lemma~(\ref{DerivChar}), $\varphi_{\rho}$ is uniformly continuous and by Laplace's method we obtain the asymptotic expression
$$ \|[\vec{X},\Gamma_{t}(\rho)]\|_{2}^{2}\sim t^{2} \frac{2}{t}(\frac{1}{\lambda_{1}+\cdots \lambda_{d} })\frac{1}{(2\pi)^{\frac{d}{2}} } \|\mathit{\Gamma}_{t}(\rho)\|_{2}^{2}, \text{ where }   \|\mathit{\Gamma}_{t}(\rho)\|_{2}^{2}=\frac{1 }{t^{d}}\frac{1}{ (\lambda_{1}\cdots \lambda_{d})^{\frac{1}{2}} } \int d\mathbf{q} |\varphi_{\rho}(\mathbf{q})|^{2}. $$
Moreover, $\varphi_{\rho}(\mathbf{q})=\Tr[e^{i\mathbf{q}\cdot \vec{K}}\rho]=\int d\mathbf{k} e^{i\mathbf{q}\cdot \mathbf{k}}\rho(\mathbf{k},\mathbf{k})$, so $\frac{1}{(2\pi)^{\frac{d}{2}} }\int d\mathbf{q} |\varphi_{\rho}(\mathbf{q})|^{2} =\int d\mathbf{k} |\rho(\mathbf{k},\mathbf{k})|^{2}$.

For the denominator we need to compute $
\frac{1}{(2\pi)^{\frac{d}{2}} }\| (t\nabla_{\mathbf{q}}+\nabla_{\mathbf{p}}-t\nabla_{\mathbf{q}}\varphi_{\rho}(0,0)-\nabla_{\mathbf{p}}\varphi_{\rho}(0,0)   )\mathit{\Gamma}_{t}^{\prime}(\varphi_{\rho})\|_{2}$.
The term $t\nabla_{\mathbf{q}}$ commutes with $\mathit{\Gamma}_{t}^{\prime}(\varphi_{\rho})$.   The terms $\nabla_{\mathbf{p}}$ and $\nabla_{\mathbf{p}}\varphi_{\rho}(0,0)$ will be of lower order.   Although when $\nabla_{\mathbf{p}}$ acts on $\mathit{\Gamma}_{t}^{\prime}(\varphi_{\rho})(0,0))$, it brings down a factor of $t\mathbf{p}$,  $|\mathbf{p}|\sim t^{-\frac{1}{2}}$.   
\begin{align*}
\frac{1}{(2\pi)^{d} }\| (t\nabla_{\mathbf{q}}-t\nabla_{\mathbf{q}}\varphi_{\rho}(0,0) )\mathit{\Gamma}_{t}^{\prime}(\varphi_{\rho})\|_{2}^{2}\rightarrow\frac{t^{2}}{(2\pi)^{d} }\int d\mathbf{q}d\mathbf{p}  e^{-t\langle \mathbf{p}| A\mathbf{p}\rangle  } |\nabla_{\mathbf{q}}\varphi_{\rho}(\mathbf{q},\mathbf{p})-  (\nabla_{\mathbf{q}}\varphi)_{\rho}(0,0)\varphi_{\rho} |^{2}.
\end{align*}
Again the Gaussian weight is on the surface $\mathbf{p}=0$.  In the limit $t\rightarrow \infty$ this is asymptotic to 
$$\frac{t^{2}t^{-\frac{d}{2}} }{(2\pi)^{\frac{d}{2}}\det(A^{k,k})^{\frac{1}{2}} } \int d\mathbf{q} | \nabla_{\mathbf{q}}\varphi_{\rho}(\mathbf{q},0)-\nabla_{\mathbf{q}}\varphi_{\rho}(0,0)\varphi_{\rho}(\mathbf{q},0)|^{2}.   $$			The integral on the right can be rewritten as $ \int d\mathbf{k}| \mathbf{k}\rho(\mathbf{k},\mathbf{k})-E[\vec{K}\rho]\rho(\mathbf{k},\mathbf{k})|^{2}$.\vspace{.5cm}

\noindent \textbf{Case 3}:\\
Since $A$ is positive definite, just the first term alone yields exponential decay for phase space points $(\mathbf{q},\mathbf{p})$ away from the origin.    By changing variables $t\mathbf{p}\rightarrow \mathbf{p}$,
 \begin{align*}
 \frac{1}{(2\pi)^{d}t^{d} }\int d\mathbf{q}d\mathbf{p}e^{-\frac{t}{2}\left\langle  \begin{tiny} \begin{pmatrix} \mathbf{q}\\ \frac{1}{t}\mathbf{p} \end{pmatrix}\end{tiny}| A \begin{tiny} \begin{pmatrix} \mathbf{q}\\ \frac{1}{t}\mathbf{p} \end{pmatrix}\end{tiny}\right\rangle -\frac{2t }{3}\left\langle \begin{tiny} \begin{pmatrix}\mathbf{p}-\frac{3}{2}\mathbf{q} \\ \frac{3}{2t}\mathbf{p} \end{pmatrix}\end{tiny}  |A \begin{tiny} \begin{pmatrix}\mathbf{p}-\frac{3}{2}\mathbf{q} \\ \frac{3}{2t}\mathbf{p} \end{pmatrix}\end{tiny}\right\rangle  } |\varphi_{\rho}(\mathbf{q},  \frac{1}{t}\mathbf{p})|^{2}.
 \end{align*}
 In the limit $t\rightarrow \infty$, the contributions from terms including $\frac{1}{t}\mathbf{p}$ become negligible for the asymptotics.   The multiplication factor of $\frac{1}{t}$ on the variable $\mathbf{p}$ in $|\varphi_{\rho}(\mathbf{q}, \frac{1}{t}\mathbf{p})|^{2}$ can only make the function more amenable to Laplace methods since the function is effectively spreading out in the $\mathbf{p}$ variable.  Neglecting these terms gives the same asymptotics as the first case.  

\end{proof}

In the next theorem we consider the case where there is also noise is also a Poisson contribution to the noise.  First we have the following lemma about classical characteristic functions.   For an origin symmetric measure positive $\nu$ on $\R^{d}$ satisfying $\int d\nu(\mathbf{k})\frac{|\mathbf{k}|^{2}}{1+|\mathbf{k}|^{2}}<\infty$, we define the function 
$$\psi_{\nu}(\mathbf{l})=\int d\nu(\mathbf{k}) (\textup{cos}(\mathbf{k}\cdot \mathbf{l})-1).$$
when $\nu(\R^{d})<\infty$, $\psi_{\nu}(\mathbf{l})=\varphi_{\nu}(\mathbf{l})-\nu(\R^{d})$.   

\begin{lemma}\label{ClassicalChar}
Let $\nu$ be a positive and possibly infinite measure on $\R^{d}$ such that $\nu$ is symmetric about the origin, and 
$$\int d\nu(\mathbf{k})|\mathbf{k}|^{2}<\infty.$$
Then the first and second derivatives are bounded and continuous and an absolute maximum occurs at the origin.  Moreover, if $B$ is the matrix of second moments $B=\int d\nu(\mathbf{k})\mathbf{k}\otimes\mathbf{k}$,    then for any $\epsilon$ there exits a $\delta$ such that for all $|\mathbf{l}|\leq \delta$
$$  -\frac{(1+\epsilon)}{2}\langle \mathbf{l}| B\mathbf{l}\rangle  -\frac{\epsilon}{2}|\mathbf{l}|^{2}\leq \psi_{\nu}(\mathbf{l}) \leq -\frac{(1-\epsilon)}{2}\langle \mathbf{l} | B\mathbf{l}\rangle +\frac{\epsilon}{2}|\mathbf{l}|^{2}. $$

 If in addition $\nu$ has a density, then the  absolute maximum  of $\psi_{\nu}$ is obtained only at the origin and  for any $\epsilon$ there exists a $\delta$ such that for all $|\mathbf{l}|\leq \delta$,
$$   -\frac{(1+\epsilon)}{2}\langle \mathbf{l}| B\mathbf{l}\rangle  \leq \psi_{\nu}(\mathbf{l}) \leq -\frac{(1-\epsilon)}{2}\langle \mathbf{l}| B\mathbf{l}\rangle .   $$
Finally $\sup_{|\mathbf{l}|=\delta}-\frac{(1-\epsilon)}{2}\langle \mathbf{l}| B\mathbf{l}\rangle >\sup_{|\mathbf{l}|\geq \epsilon }\psi_{\nu}(\mathbf{l})$.

\end{lemma}

\begin{proof} \text{ }
We can rewrite the expression for $\psi_{\nu}$ as:
\begin{align}\label{Psi}
\psi_{\nu}(\mathbf{l})=\int d\mu(\mathbf{k}) |\mathbf{k}|^{2}\Big(\frac{\textup{cos}(\mathbf{k}\cdot \mathbf{l})-1}{|\mathbf{k}|^{2}}\Big).
\end{align}
The first and second derivatives in $\mathbf{l}$ of the family of functions $f_{\mathbf{k}}(\mathbf{l})$, where
$$f_{\mathbf{k}}(\mathbf{l})=\frac{\textup{cos}(\mathbf{k}\cdot \mathbf{l})-1}{|\mathbf{k}|^{2}},$$
are continuous and  uniformly bounded.   By  our assumption on $\nu$, the measure defined by $d\nu(\mathbf{k})|\mathbf{k}|^{2}$ has finite total mass.   It follows that $\psi_{\nu}$ is bounded with bounded and continuous first and second derivatives.   $\psi_{\nu}$ is real, centrally symmetric, and the first derivatives of $\psi_{\nu}$ are zero at the origin.    Thus $\phi_{\nu}$ obtains a absolute maximum at the origin.    If $\nu$ has a density $\frac{d\nu}{d\mathbf{k}}$ then the absolute maximum is unique since this is the only time point $\mathbf{l}$ at which all the phases in the integral~(\ref{Psi}) are aligned.  

$D^{2}\psi_{\nu}(\mathbf{l})$  can be expressed according to the formula:
$$D^{2}\psi_{\nu}(\mathbf{l})=-\int d\nu(\mathbf{k})\mathbf{k}\otimes \mathbf{k} e^{i\mathbf{k}\cdot \mathbf{l}}.$$
 For $\mathbf{l}=0$, this expression is equal to $-B$.  $D^{2}\psi_{\nu}(\mathbf{l})$ is continuous in the operator norm since its components are continuous and all norms are equivalent over finite dimensional spaces.    

In the direction $\mathbf{l}$, we can write the second order Taylor expansion:
\begin{align*}\label{SecDer}
\psi_{\nu}(\mathbf{l})=\psi_{\nu}(0)+\nabla \psi_{\nu}(0)\mathbf{l}+\int_{0}^{1}ds\int_{0}^{s}dr\langle \mathbf{l}|D^{2}\psi_{\nu}(r\mathbf{l})\mathbf{l}\rangle=  \int_{0}^{1}ds\int_{0}^{s}dr\langle \mathbf{l}| D^{2}\psi_{\nu}(r\mathbf{l})\mathbf{l}\rangle 
\end{align*}
Since $-D^{2}\psi_{\nu}(\mathbf{l})$ is continuous with respect to the operator norm and  positive semidefinite at zero, it follows for any $\epsilon$ there exists a $\delta$ such that
$$     (1-\epsilon)B-\epsilon I_{d}  \leq  -D^{2}\psi_{\nu}(\mathbf{l})  \leq (1+\epsilon)B +\epsilon I_{d} $$
for all $|\mathbf{l}|\leq \epsilon$.   Applying this inequality to the formula~(\ref{SecDer}), we have
$$-\frac{1}{2}(1+\epsilon)\langle \mathbf{l}|B\mathbf{l}\rangle-\frac{\epsilon}{2}|\mathbf{l}|^{2} \leq \psi_{\nu}(\mathbf{l})\leq -\frac{1}{2}(1-\epsilon)\langle \mathbf{l} | B\mathbf{l}\rangle  +\frac{\epsilon}{2}|\mathbf{l}|^{2}.$$

In the case where $\nu$ has a density, then the matrix $B$ is positive definite since the integration of terms $\mathbf{k}^{2} $ cannot have its support over some lower dimensional space.   By continuity of  $D^{2}\psi_{\nu}(\mathbf{l})$, for any $\epsilon$ we can pick a $\delta$ such that
$$-\frac{1}{2}(1+\epsilon)\langle \mathbf{l}| B\mathbf{l}\rangle \leq \psi_{\nu}(\mathbf{l})\leq -\frac{1}{2}(1-\epsilon)\langle \mathbf{l} |B\mathbf{l}\rangle.$$

Furthermore, we can choose a $\delta$ small enough such that $\psi_{\nu}(\mathbf{l})$ is concave down for all $|\mathbf{l}|<\delta$ (and hence decreasing radially from the origin), and such that any local maximum that is not the origin is less than $\inf_{|\mathbf{l}|\leq \delta} \varphi_{\nu}(\mathbf{l})$.   Hence 
$$\sup_{|\mathbf{l}|\geq \delta} \psi_{\nu}(\mathbf{l})= \sup_{|\mathbf{l}|= \delta} \psi_{\nu}(\mathbf{l})\leq   \sup_{|\mathbf{l}|= \delta} -\frac{1}{2} (1-\epsilon)\langle \mathbf{l}| B\mathbf{l}\rangle .$$

\end{proof}

For the L\'evy measure $\mu$ define
$$B=\int d\mu(\mathbf{x},\mathbf{k}) \begin{tiny}\begin{pmatrix} \mathbf{x} \\ \mathbf{k} \end{pmatrix}\end{tiny}\otimes \begin{tiny}\begin{pmatrix} \mathbf{x} \\ \mathbf{k} \end{pmatrix}\end{tiny}=\begin{tiny} \begin{pmatrix} B^{x,x} & B^{x,k}\\ B^{k,x} & B^{k,k} \end{pmatrix}\end{tiny} $$.

\begin{theorem}\label{ThmPoisson}
Let $\rho$ be a density operator such that $J \rho \in \TC_{1}(L^{2}(\R^{d}))$ for all \\ $J\in \{X_{1},\cdots, X_{d},K_{1},\cdots, K_{d}\}$.  Let $\varphi_{\Gamma_{t}}(\rho)$ satisfy Equation~(\ref{LevyKhinchin}), where $\mu$ has origin symmetry $\mu(\mathbf{x},\mathbf{k})=\mu(-\mathbf{x},-\mathbf{k})$ and the second moments
 $$\int d\nu(\mathbf{x},\mathbf{k})(|\mathbf{x}|^{2}+|\mathbf{k}|^{2})<\infty.$$
\begin{enumerate}
\item 
 With $\mu=\delta(\mathbf{x}) \nu(\mathbf{k}) $ and  $A^{k,x}=A^{x,k}=A^{k,k}=0$, then if $\nu$ has a density or $A^{x,x}$ is positive definite, then we have the same asymptotics as (1) from~(\ref{ThmBrown}) with $A^{x,x}$ replaced by $A^{x,x}+B^{x,x}$. 
 \item
With $\mu= \nu(\mathbf{x})\delta(\mathbf{k})$ and $A^{x,x}=A^{k,x}=A^{x,k}=0$,  then if $\nu$ has a density or $A^{k,k}$ is positive definite, then we have the same asymptotics as (2) from~(\ref{ThmBrown}) with $A^{x,x}$ replaced by $A^{x,x}+B^{x,x}$. 
\item 
 If $\mu$ has a density or $A$ is positive definite, then we have the same asymptotics as (3) from~(\ref{ThmBrown}) with $A$ replaced by $A+B$.

\end{enumerate}

\end{theorem}

\begin{proof}
The basic idea of this proof is that for long time periods we can effectively approximate the exponent of the expression $\mathit{\Gamma}_{t}^{\prime}$, $\int_{0}^{t}ds \psi_{\mu}(\mathbf{q}-s\mathbf{p},\vec{p})$ as a quadratic through a Taylor expansion of $\psi_{mu}$ at zero.   Once we have shown this, then we can refer to our results from proposition~(\ref{ThmBrown}).  

 Just as in~(\ref{ThmBrown}), to approximate $\|[\vec{X},\Gamma_{t}(\rho)]\|_{2}$ we need to handle 
$$\frac{1}{(2\pi)^{d}}\int d\mathbf{p}d\mathbf{q}|\mathbf{q}-t\mathbf{p}|^{2}e^{-\int_{0}^{t}ds \left\langle \begin{tiny}\begin{pmatrix} \mathbf{q}-s\mathbf{p} \\ \mathbf{p} \end{pmatrix}\end{tiny}| A\begin{tiny}\begin{pmatrix} \mathbf{q}-s\mathbf{p}  \\ \mathbf{p} \end{pmatrix}\end{tiny}\right\rangle -2\int_{0}^{t}ds\psi_{\mu}( \mathbf{q}-s\mathbf{p}, \mathbf{p}) }|\varphi_{\rho}(\mathbf{q},\mathbf{p})|^{2}.$$.
We will show that outside of some small ball around the origin, all phase space points are experiencing a uniform super-polynomial decay.  

\noindent \textbf{Case 1}:
By~(\ref{ClassicalChar}), for any $\epsilon$ there exist a $\delta$ such that $|\mathbf{l}|\leq \delta $
$$-\frac{1}{2}(1+\epsilon)\langle \mathbf{l}| B^{x,x}\mathbf{l}\rangle  \leq \psi_{\nu}(\mathbf{l})\leq -\frac{1}{2}(1-\epsilon)\langle \mathbf{l}| B^{x,x}\mathbf{l}\rangle , $$
and
$$\sup_{|\mathbf{l}|\geq \delta} \psi_{\nu}(\mathbf{l}) \leq    \sup_{|\mathbf{l}|= \delta} -\frac{1}{2}(1-\epsilon)\langle \mathbf{l}| B^{x,x} \mathbf{l}\rangle .$$  
Define the constant $d$,  $d=\sup_{|\mathbf{l}|= \frac{\delta}{3} } -(1-\epsilon)\langle\mathbf{l}|B\mathbf{l}\rangle $.   Define $S_{\frac{\delta}{3},t}$ to be the set of phase space points $(\mathbf{q},\mathbf{p})$ such that $|\mathbf{q}-s\mathbf{p}|>\frac{\delta}{3}$ for at least a fraction of $\frac{1}{\sqrt{t}}$ of intermediate times $s$ in the interval $[0,t]$.  Up to time $t$ these points have a maximum decay factor of $e^{d\sqrt{t}}$.   It follows that for large times $t$ these points have a super-polynomial  and thus negligible contribution.   On the other hand, points in $S_{\frac{\delta}{3},t}^{c}$ satisfy $|\mathbf{q}-s\mathbf{p}|< \delta$ for all intermediary times $s\in [0,t]$ as long as $t>4$.   This follows since the moving point $\mathbf{q}-s\mathbf{p}$ requires a time interval of at least length $t-\sqrt{t}$ to  travel through an arc of $B_{\frac{\delta}{3}}(0)$.  Hence in an additional time period of length $\sqrt{t}$, it can not travel the minimum distance $\frac{2\delta}{3}$ required to escape the $\delta$ ball as long as $t>4$.  It follows that for all points in $S_{\frac{\delta}{3},t}^{c}$ and for all intermediary times $s$ we have
$$-\frac{1}{2}(1+\epsilon)\langle \mathbf{q}-s\mathbf{p}| B^{x,x}(\mathbf{q}-s\mathbf{p})\rangle  \leq \psi_{\nu}(\mathbf{q}-s\mathbf{p})\leq -\frac{1}{2}(1-\epsilon)\langle \mathbf{q}-s\mathbf{p}| B^{x,x}(\mathbf{q}-s\mathbf{p})\rangle .$$
The region of points in $S_{\frac{\delta}{2},t}$ is negligible so we have the asymptotic upper and lower bounds $\mp$ for our original expression as
$$\frac{1}{(2\pi)^{d}}\int d\mathbf{p}d\mathbf{q}|\mathbf{q}-t\mathbf{p}|^{2}e^{-\int_{0}^{t}ds\langle \mathbf{q}-s\mathbf{p}| A^{x,x}(\mathbf{q}-s\mathbf{p})\rangle -\int_{0}^{t}ds (1\mp\epsilon)\langle \mathbf{q}-s\mathbf{p}| B^{x,x}(\mathbf{q}-s\mathbf{p})\rangle  }. $$
By applying our results from~(\ref{ThmBrown}) with $A^{x,x}$ replaced by $A^{x,x}+(1\pm \epsilon)B^{x,x}$ and letting $\epsilon$ go to zero we get our asymptotics.    

Now we deal with the case where $A^{x,x}$ is positive definite, but $\nu$ is not assumed to have a density.   Given any $\delta$ the contribution from points in $S_{\frac{\delta}{2},t}^{c}$ will have a negligible effect on the decay rate by the same argument as above through the term 
$$-\int_{0}^{t}ds \langle \mathbf{q}-s\mathbf{p}|A^{x,x}(\mathbf{q}-s\mathbf{p})\rangle .$$
Since points in $S_{\frac{\delta}{2},t}$ have that $|\vec{q}-s\vec{p}|<\delta$ for all intermediate times and by~(\ref{ClassicalChar}), for any $\epsilon$ there is a $\delta$ such that  $2\int_{0}^{t}ds\psi_{\nu}(\mathbf{q}-s\mathbf{p})$ is bounded above and below by
$$
-\int_{0}^{t}ds \big( (1\mp \epsilon)\langle \mathbf{q}-s\mathbf{p}|B^{x,x}(\mathbf{q}-s\mathbf{p}) \rangle \pm \epsilon |\mathbf{q}-s\mathbf{p}|^{2}\big). $$
By taking $\epsilon$ less than the smallest eigenvalue of $A^{x,x}$, we can apply (1) of~(\ref{ThmBrown}), and take the limit as $\epsilon$ goes to zero to get the asymptotics.    

The other cases are handled similarly.

\end{proof}

Now we give a theorem characterizing the asymptotic form of $\Gamma_{t}(\rho)$.   It doesn't require any new ideas as opposed to those appearing in the above proves.  

\begin{theorem}\label{Relaxation}
Let  the dynamics $\Gamma_{t}$ satisfy the conditions of~(\ref{ThmPoisson}) for cases $1$ or $3$ and $\tilde{\rho}_{t}$ satisfy~(\ref{Rel}), then 
\begin{eqnarray}\label{Relax}
\frac{\|\Gamma_{t}(\rho)-\tilde{\rho}_{t}\|_{2}}{ \|\tilde{\rho}_{t}\|_{2}}\rightarrow 0.
\end{eqnarray}

\end{theorem}

\begin{proof}
By using~(\ref{Isometry}), we can rewrite~(\ref{Relax}) using the quantum characteristic functions $\varphi_{\tilde{\rho}_{t}}$ and $\varphi_{\Gamma_{t}(\rho)}$.   We can then apply the same techniques as  in~(\ref{ThmPoisson}) to replace the Poisson noise by bounds using Gaussian noise.   By Laplace methods, we can see that the numerator tends to zero faster than the denominator. 
\end{proof}

\begin{theorem}\label{Classicality}
Let  the dynamics $\Gamma_{t}$ satisfy the conditions of~(\ref{ThmPoisson}) for cases $1$ or $3$ and
$p_{t}(\mathbf{x},\mathbf{v})$ be the probability distribution for the classical L\'evy process at time $t$, then  
$$\frac{\big(\int d\mathbf{x}d\mathbf{v}| \mathcal{W}_{\Gamma_{t}(\rho)}(\mathbf{x},\mathbf{v})-p_{t}(\mathbf{x},\mathbf{v})|^{2} \big)^{\frac{1}{2}}}{\big(\int d\mathbf{x}d\mathbf{v}|p_{t}(\mathbf{x},\mathbf{v})|^{2} \big)^{\frac{1}{2}}}\rightarrow 0 ,    $$
where $\mathcal{W}_{\Gamma_{t}(\rho)}(\mathbf{x},\mathbf{v})$ is the Wigner distribution for the state $\Gamma_{t}(\rho)$.
\end{theorem}
\begin{proof}
Taking the Fourier transform of the integrand in the numerator we are left with the difference between the quantum characteristic function of $\Gamma_{t}(\rho)$ and the characteristic function of $p_{t}(\mathbf{x},\mathbf{v})$.   The multiplication factor governing the noise in the formula for $\varphi_{\Gamma_{t}(\rho)}(\mathbf{q},\mathbf{p})$ causes points away from the origin to vanish.     However, $\varphi_{\rho}(\mathbf{q}+t\mathbf{p},\mathbf{p})$ is continuous and takes the value one at the origin, and this can be effectively replaced by the function that is $1$ everywhere.     
\end{proof}

\newpage
\appendix

\begin{center}
    {\bf \Large APPENDIX \normalsize }
  \end{center}

\section{The Quantum Characteristic Function }
  The quantum characteristic function is defined as $\varphi_{\rho}(\mathbf{q},\mathbf{p})=\Tr[W_{\mathbf{q},\mathbf{p}}\rho]$ for $\rho\in \TC_{1}(\R^{d})$.  Weyl operators satisfy the multiplication formula
\begin{eqnarray}\label{NotBCH}
  W_{(\mathbf{q}_{1},\mathbf{p}_{1})}W_{(\mathbf{q}_{2},\mathbf{p}_{2})}=e^{\frac{i}{2}(-\mathbf{q}_{1}\cdot \mathbf{p}_{2}+\mathbf{p}_{1}\cdot \mathbf{q}_{2})}    W_{(\mathbf{q}_{1}+\mathbf{q}_{2},\mathbf{p}_{1}+\mathbf{p}_{2})} 
\end{eqnarray}
   Formally, this formula follows from the Baker-Campbell-Hausdorf (BCH) formula.   Using~(\ref{NotBCH}) with the characteristic function formula 
\begin{align}e^{\frac{i}{2}\mathbf{q}\cdot \mathbf{p}}\Tr[e^{i\mathbf{q}\cdot \vec{K}}e^{i\mathbf{p}\cdot \vec{X}}\rho].
\end{align} 

Since $e^{i\mathbf{q}\cdot \vec{K}}$ acts as a translation operator by $\mathbf{q}$ in the $x$-basis, intuitively we can apply the formula for a trace to reach the equality
$$e^{\frac{i}{2}\mathbf{q}\cdot \mathbf{p}}\Tr[e^{i\mathbf{q}\cdot \vec{K}}e^{i\mathbf{p}\cdot \vec{X}}\rho]=e^{-\frac{i}{2}\mathbf{q}\cdot \mathbf{p}} \int d\mathbf{x} e^{i\mathbf{p}\cdot \mathbf{x}}\rho(\mathbf{x}-\mathbf{q},\mathbf{x}).  $$
Now taking the Fourier transform in the $\mathbf{p}$ variable we get 
$$\frac{1}{(2\pi)^{d} } \int d\mathbf{p}e^{-i\mathbf{x}\cdot \mathbf{p}}\varphi_{\rho}(\mathbf{q},\mathbf{p})=\rho(\mathbf{x}-\frac{q}{2},\mathbf{x}+\frac{q}{2}).$$
The Fourier transform of the $\mathbf{q}$ variable is by definition the Wigner distribution function $\mathcal{W}_{\rho}(\mathbf{x},\mathbf{v})$.     Hence the quantum characteristic function and Wigner distribution function are related by a Fourier transform in both variables:
$$ \frac{1}{(2\pi)^{2d} }\int d\mathbf{q}d\mathbf{p}e^{-i\mathbf{p}\cdot \mathbf{x}-i\mathbf{q}\cdot \mathbf{v}}\varphi_{\rho}(\mathbf{q},\mathbf{p})=\mathcal{W}_{\rho}(\mathbf{x},\mathbf{v}).$$

\begin{lemma}\label{Use}
Suppose $\rho$ be a density operator, and $J \rho  \in \TC_{2}(L^{2}(\R^{d}))$ for \\$G\in \{X_{1},\cdots,X_{d}, K_{1}, \cdots K_{d} \}$.  Let $\mathbf{q}_{0},\mathbf{p}_{0}\in \R^{d}$, with $|\mathbf{q}_{0}|^{2}+|\mathbf{p}_{0}|^{2}=1$.  Then we have that
$$ h^{-1}(W_{h(\mathbf{q}_{0},\mathbf{p}_{0})}-I  )\rho \rightarrow i(\mathbf{q}_{0}\cdot \vec{X}+\mathbf{p}_{0}\cdot \vec{K})\rho \text{, and}$$ 
$$  \rho (  W_{h(\mathbf{q}_{0},\mathbf{p}_{0}) }-I )h^{-1} \rightarrow i\rho(\mathbf{q}_{0}\cdot \vec{X}+\mathbf{p}_{0}\cdot \vec{K}),   $$
 where the convergence is in the trace norm.  

\end{lemma}

\begin{proof}
Define the self-adjoint operator $H= \mathbf{q}_{0}\cdot \vec{X}+\mathbf{p}_{0}\cdot \vec{K}$ so we can write $W_{h(\mathbf{q}_{0},\mathbf{p}_{0}) }=e^{ihH}$.   By our conditions on $\rho$ and the triangle inequality, $H\rho$ is trace-class.   Technically, $H\rho$ is defined as the bounded operator (traceclass even) determining the bilinear form $\mathit{B}(g,f)=\langle H g| \rho f\rangle$, for $g\in \textup{D}(H)$ and $f\in L^{2}(\R^{d})$.     In particular, the boundedness of the $\mathit{B}$ implies that $\rho$ maps arbitrary elements in $L^{2}(\R^{d})$ to $D(H)$.  

 Note that $|h^{-1}(W_{(h\mathbf{q}_{0},h\mathbf{p}_{0})}-I)|\leq |H|$.   However, for two operators $A$, $B$ such that $0\leq A\leq B$,  then $\rho A^{2} \rho\leq \rho B^{2}\rho $ and $\| A\rho \|_{1}\leq \|B\rho\|_{1}$.   To see that $\| A\rho \|_{1}\leq \|B\rho\|_{1}$, let $g_{j}$ an orthonormal basis of eigenvectors for $\rho B^{2}\rho$, then we have
$$\|A\rho\|_{1}\leq \sum_{j}(\langle g_{j}| \rho A^{2} \rho g_{j}\rangle )^{\frac{1}{2}}\leq \sum_{j}(\langle g_{j}| \rho B^{2} \rho g_{j}\rangle )^{\frac{1}{2}}= \sum_{j}\langle g_{j}|( \rho B^{2} \rho  )^{\frac{1}{2}}g_{j}\rangle= \|B\rho\|_{1},$$
where the first inequality above follows by writing $\rho A^{2}\rho$ in terms of its spectral decomposition and applying Jensen's inequality.   Applying this fact with $A=h^{-1}(W_{(h\mathbf{q}_{0},h\mathbf{p}_{0})}-I)$ and $B=H$, we have that
$$\|h^{-1}(e^{ihH}-I)\rho \|_{1}\leq  \| H \rho\|_{1}.     $$
Hence, $h^{-1}(e^{ihH}-I)\rho$ is trace class.  By the singular value decomposition, there exists a sequence of finite dimensional projections $P_{n}$ such that $H\rho P_{n}$ converges to $H\rho$ in the trace norm.    
\begin{multline}
\|( h^{-1}(e^{ihH}-I)-iH)\rho\|_{1}\leq \| ( h^{-1}(e^{ihH}-I)-iH)\rho P_{n}\|_{1}\\ +\| ( h^{-1}(e^{ihH}-I)-iH)\rho (I-P_{n})\|_{1}
\end{multline}
The second term is bounded by $2\|H\rho(I-P_{n})\|_{1}$ and we can pick a $n$ large enough so that  this term is smaller than $\frac{\epsilon}{2}$.    On the other hand, the image of $\rho P_{n}$ is finite dimensional and contained in the domain of $H$.   Using Stone's Theorem~\cite{Reed} over that finite dimensional space, we can pick an $h$ such that 
$$\|( h^{-1}(e^{ihH}-I)-iH)\rho P_{n}\|_{\infty}< \frac{\epsilon}{2n}, \text{ and hence } \|( h^{-1}(e^{ihH}-I)-iH)\rho P_{n}\|_{\infty}< \frac{\epsilon}{2}. $$  
Hence we have the trace norm convergence
$$ h^{-1}(W_{h(\mathbf{q}_{0},\mathbf{p}_{0})}-I  )\rho \rightarrow iH\rho. $$

Similarly $\rho h^{-1}(W_{(h\mathbf{q}_{0},h\mathbf{p}_{0})}-I)\rightarrow i\rho H$.   

\end{proof}

\begin{lemma}\label{DerivChar}
Suppose $\rho$ be as density operator and  $J\rho \in \TC_{1}(L^{2}(\R^{d}))$ for \\ $G\in \{X_{1},\cdots,X_{d}, K_{1}, \cdots K_{d} \}$.   It follows that the first derivatives of $\varphi_{\rho}(\mathbf{q},\mathbf{p})$ are bounded and continuous.   Moreover, for $\mathbf{q}_{0},\mathbf{p}_{0}\in \R^{d}$ we have the formula:
$$  \begin{pmatrix} \mathbf{q}_{0} \\ \mathbf{p}_{0} \end{pmatrix}\cdot \nabla_{ ( \mathbf{q}, \mathbf{p} )      } \varphi_{\rho}(\mathbf{q},\mathbf{p})= i\varphi_{ \{\mathbf{q}_{0}\cdot \vec{K}+\mathbf{p}_{0}\cdot \vec{X}, \rho\} }(\mathbf{p},\mathbf{q}). $$

\end{lemma}
\begin{proof}
Let $|\mathbf{q}_{0}|^{2}+|\mathbf{p}_{0}|^{2}=1$, $h>0$, and $W_{(\mathbf{x},\mathbf{k})}=e^{i\mathbf{x}\vec{K}+i\mathbf{k}\vec{X}}$ be the Weyl operator for a translation by $(\mathbf{x},\mathbf{k})$ in phase space.  By the cyclicity of trace and action of Weyl operators
$$\varphi_{\rho}(\mathbf{q}+h\mathbf{q}_{0},\mathbf{p}+h\mathbf{p}_{0})=\Tr[W_{(\mathbf{q},\mathbf{p})} W_{(h\mathbf{q}_{0},h\mathbf{p}_{0})} \rho W_{(h\mathbf{q}_{0},h\mathbf{p}_{0})}].$$
We can write
\begin{multline}\label{bound}
\frac{1}{h}(\varphi_{\rho}(\mathbf{q}+h\mathbf{q}_{0},\mathbf{p}+h\mathbf{p}_{0})-\varphi_{\rho}(\mathbf{q},\mathbf{p}) )=\Tr[W_{(\mathbf{q},\mathbf{p})} h^{-1}(W_{(h\mathbf{q}_{0},h\mathbf{p}_{0})}-I) \rho W_{(h\mathbf{q}_{0},h\mathbf{p}_{0})}]\\ +\Tr[W_{(\mathbf{q},\mathbf{p})} \rho ( W_{(h\mathbf{q}_{0},h\mathbf{p}_{0})}-I)h^{-1} ].
\end{multline}
By Lemma~(\ref{Use}), $h^{-1}(W_{(h\mathbf{q}_{0},h\mathbf{p}_{0})}-I) \rho  $ and $\rho (W_{(h\mathbf{q}_{0},h\mathbf{p}_{0})}-I)h^{-1}$ converge to $i(\mathbf{q}_{0}\cdot \vec{K}+\mathbf{p}_{0}\cdot \vec{X})\rho$ and $i \rho(\mathbf{q}_{0}\cdot \vec{K}+\mathbf{p}_{0}\cdot \vec{X})$, respectively, in the $1$-norm.   Since $W_{(\mathbf{q},\mathbf{p})} $ and  $W_{(h\mathbf{q}_{0},h\mathbf{p}_{0})}$ are unitary, they are bounded in the operator norm, and the above expression converges to    
\begin{align*}
i\Tr[W_{(\mathbf{q},\mathbf{p})}  (\mathbf{q}_{0}\cdot \vec{K}+\mathbf{p}_{0}\cdot \vec{X}) \rho] +i\Tr[W_{(\mathbf{q},\mathbf{p})}  \rho (\mathbf{q}_{0}\cdot \vec{K}+\mathbf{p}_{0}\cdot \vec{X})   ].
\end{align*}
Since $ (\mathbf{q}_{0}\cdot \vec{K}+\mathbf{p}_{0}\cdot \vec{X}) \rho$ and $\rho (\mathbf{q}_{0}\cdot \vec{K}+\mathbf{p}_{0}\cdot \vec{X}) $ are trace class,  the expression above is bounded and continuous.  
Moreover, it can be written as
$$  \begin{pmatrix} \mathbf{q}_{0} \\ \mathbf{p}_{0} \end{pmatrix}\cdot \nabla_{ ( \mathbf{q}, \mathbf{p} )      } \varphi_{\rho}(\mathbf{q},\mathbf{p})= i\varphi_{ \{\mathbf{q}_{0}\cdot \vec{K}+\mathbf{p}_{0}\cdot \vec{X}, \rho\} }(\mathbf{p},\mathbf{q}). $$

\end{proof}

\begin{proposition}\label{Isometry}
Consider the complex Hilbert Space $\Hi=L^{2}(\R^{d})$.   The map $\mathcal{C}$ sending trace-class operators $\rho$ to their quantum characteristic functions $\varphi_{\rho}$ extends to an isometry from the Hilbert-Schmidt class operators $\TC_{2}(L^{2}(\R^{d}))$ to $L^{2}(\R^{d}\times \R^{d}, \frac{1}{(2\pi)^{d}}d\mathbf{q}d\mathbf{p} )$.
\end{proposition}

The above proposition can be proven be considering that $L^{2}(\R^{d})=\Gamma(\C^{d})$, where $\Gamma(\C^{d})$ is the Fock space generated by $\C^{d}$.   $\varphi_{\rho}$ is explicitly computable in the case that $\rho= |e(v)\rangle \langle e(u)|$ where $e(u)$ and $e(v)$ are exponential vectors.  It can then be shown that
$$\Tr[\rho_{1}^{*}\rho_{2}]=\frac{1}{(2\pi)^{d}}\int d\mathbf{x}\mathbf{p}\bar{\varphi}_{\rho_{1}}(\mathbf{x},\mathbf{p})\varphi_{\rho_{2}}(\mathbf{x},\mathbf{p}),    $$
where $\rho_{1}$ and $\rho_{2}$ are two non-orthogonal projections constructed with exponential vectors.   This property can then be extended Hilbert-Schmidt operators in general.

\section*{Acknowledgments}
The author would like to thank Bruno Nachtergaele for valuable discussions at the outset of this work  and generously providing suggestions towards the presentation and organization of this article.   Partial financial support for my work has come from Graduate Student Research (GSR) fellowships  funded by the National Science Foundation (NSF  \# DMS-0303316 and DMS-0605342).


\begin{thebibliography}{99}
\bibitem{Alicki}R. Alicki:  \emph{A Search for a Border Between Classical and Quantum Worlds} Phys. Rev. \textbf{A65}, 034104 (2002).
\bibitem{Dyn} R. Alicki, M. Fannes:  \emph{Quantum Dynamical Systems}, Oxford University Press, 2001.
\bibitem{Semi}R. Alicki, K. Lendi:  \emph{Quantum Dynamical Semigroups and Applications}.  Springer-Verlag, Berlin, 1987.
\bibitem{Feller} D. Applebaum: \emph{L\'evy Processes and Stochastic Calculus}, Cambridge University Press,  2004.
\bibitem{Caldeira} A. O. Caldeira, A. J. Leggett:  \emph{Influence of Damping on Quantum Interference: An Exactly Solvable Model},  Phys. Rev. \textbf{A31}, 1059--1066 (1985).
\bibitem{Cheb}A. M. Chebotarev, F. Fagnola: \emph{Sufficient Conditions for Conservativity of Minimal Quantum Dynamical Semigroups}, J. Funct. Anal. \textbf{153}, 382--404 (1998).
\bibitem{MME}E. B. Davies: \emph{Markovian Master Equations}, Comm. Math. Phys. \textbf{39}, 91--110 (1974).
\bibitem{Folland}G. B. Folland:  \emph{Real Analysis},  John Wiley and Sons, 1999.  
\bibitem{ESL}M. R. Gallis, G. N. Fleming: \emph{Environmental and Spontaneous Localization},  Phys. Rev. \textbf{A42}, 38-48 (1990).
\bibitem{Ghirardi}G. C. Ghirardi, A. Rhimini, T. Weber:  \emph{Unified Dynamics for Microscopic and Macroscopic Systems},  Phys. Rev.  \textbf{D34}, 470--491 (1986).
\bibitem{Talbot}L. Hackerm\"uller, K. Hornberger, B. Brezger, A. Zeilinger, M. Arndt:  \emph{Decoherence in a Talbot-Lau Interferometer: the Influence of Molecular Scattering}, Appl. Phys.  \textbf{B77}, 781--787 (2003).
\bibitem{Holevo} A. S. Holevo: \emph{On Conservativity of Covariant Dynamical Semigroups},  Rep. Math. Phys. \textbf{33}, 95--110 (1993).
\bibitem{Dissipate} A. S. Holevo: \emph{On Dissipative Stochastic Equations in a Hilbert Space}, Probab. Theory Related Fields 104, 483-500 (1996). 
\bibitem{Cov}A. S. Holevo: \emph{Covariant Quantum Markovian Evolutions}, J. Math. Phys. \textbf{37}, 1812--1832  (1996).
\bibitem{Unbounded} A. S. Holevo: \emph{Covariant Quantum Dynamical Semigroups: Unbounded Generators},  67--81, Lecture Notes in Phys., 504, Springer, Berlin, (1998).
\bibitem{Sipe} K. Hornberger, J. Sipe:  \emph{Collisional Decoherence Reexamined},  Phys. Rev. \textbf{A68}, 012105 (2003).
\bibitem{Exper}K. Hornberger, S. Uttenthaler, B. Brezger, L. Hackerm\"uller, M. Arndt, and A. Zeilinger: \emph{Collisional Decoherence Observed in Matter Wave Interferometry}, Phys. Rev. Lett. \textbf{90} 160401 (2003).
\bibitem{Kato} T. Kato: \emph{Perturbation Theory for Linear Operators}, 2nd edition.   Springer-Verlag, 1984.
\bibitem{Joos}E. Joos, H. D. Zeh: \emph{The Emergence of Classical Properties Through Interaction with the Environment}, Z. Phys.  \textbf{B59}, 223--243 (1985).
\bibitem{Decoh}E. Joos, H. D. Zeh, C. Kiefer, D. Guilini, J. Kupsch, I. \-O. Stamatescu:  \emph{Decoherence and the Appearance of a Classical World in Quantum Theory}, Springer-Verlag, 2003. 
\bibitem{Lindblad}G. Lindblad: \emph{On the Generators of Quantum Dynamical Semigroups}, Comm. Math. Phys. \textbf{48}, 119--130 (1976).
\bibitem{Lutz} E. Lutz: \emph{Anomalous L\'evy decoherence}, Phys. Lett. A \textbf{293}, 123-128 (2002).  
\bibitem{Morikawa}M. Morikawa:  \emph{Quantum Decoherence and Classical Correlation in Quantum Mechanics}, Phys. Rev.  \textbf{D42}, 2929--2931 (1990).
\bibitem{Reed} M. Reed, B. Simon: \emph{Functional Analysis},  Academic Press, 1980.
\bibitem{Vacchini0}B. Vacchini: \emph{Completely Positive Quantum Dissipation},  Phys. Rev. Lett. \textbf{84}, 1374--1377 (2000). 
\bibitem{Vacchini}B. Vacchini: \emph{Master-Equations for the Study of Decoherence}, Int. Journ. Theor. Phys. \textbf{44}, 1011--1021 (2005).
\bibitem{Levy}B. Vacchini: \emph{Theory of Decoherence due Scattering Events and L\'evy Processes}, Phys. Rev. Lett. \textbf{95}, 230402 (2005).  
\end{thebibliography}
\end{document}